\documentclass[superscriptaddress,nofootinbib,11pt,preprintnumbers,notitlepage]{revtex4-1}
\usepackage{epsfig,amsmath,amssymb}
\usepackage{color}
\usepackage{float}
\usepackage[colorlinks]{hyperref}
\usepackage{subfigure}
\usepackage[utf8]{inputenc}
\usepackage{booktabs}
\usepackage{multirow}
\usepackage{longtable}
\usepackage{booktabs}
\usepackage{array}
\usepackage{tabularx}
\usepackage{rotating}
\usepackage{array}
\usepackage{bm,wasysym}
\usepackage{booktabs}
\usepackage{soul}

\begin{document}

\renewcommand{\(}{\left(}
\renewcommand{\)}{\right)}
\renewcommand{\{}{\left\lbrace}
\renewcommand{\}}{\right\rbrace}
\renewcommand{\[}{\left\lbrack}
\renewcommand{\]}{\right\rbrack}
\renewcommand{\Re}[1]{\mathrm{Re}\!\{#1\}}
\renewcommand{\Im}[1]{\mathrm{Im}\!\{#1\}}
\newcommand{\dd}[1][{}]{\mathrm{d}^{#1}\!\!\;}
\newcommand{\del}{\partial}
\newcommand{\nn}{\nonumber}
\newcommand{\ie}{i.e.\,}
\newcommand{\cf}{cf.\,}
\newcommand{\refeq}[1]{Eq.~(\ref{eq:#1})}
\newcommand{\refeqs}[2]{Eqs.~(\ref{eq:#1})-(\ref{eq:#2})}
\newcommand{\reffig}[1]{Fig.~\ref{fig:#1}}
\newcommand{\refsec}[1]{Section \ref{sec:#1}}
\newcommand{\reftab}[1]{Table \ref{tab:#1}}
\newcommand{\order}[1]{\mathcal{O}\({#1}\)}
\newcommand{\fv}[1]{\left(\begin{array}{c}#1\end{array}\right)}%

\def\mK{{m_{K_2^\ast}}}
\def\mKK{{m^2_{K^\ast_2}}}
\def\EK{{E_{K_2^\ast}}}

\def\tcb#1{\textcolor{blue}{#1}}
\def\tcr#1{\textcolor{red}{#1}}
\def\tcg#1{\textcolor{green}{#1}}
\def\tcc#1{\textcolor{cyan}{#1}}
\def\tcv#1{\textcolor{violet}{#1}}
\def\tcm#1{\textcolor{magenta}{#1}}
\def\tcpn#1{\textcolor{pink}{#1}}
\def\tcpr#1{\textcolor{purple}{#1}}
\definecolor{schrift}{RGB}{120,0,0}

\def \azeL{{H_0^L}}
\def \azeR{{H_0^R}}
\def \apaL{{H_\parallel^L}}
\def \apaR{{H_\parallel^R}}
\def \apeL{{H_\perp^L}}
\def \apeR{{H_\perp^R}}

%% Physics
\newcommand{\alphas}{\alpha_\mathrm{s}}
\newcommand{\alphae}{\alpha_\mathrm{e}}
\newcommand{\gfermi}{G_\mathrm{F}}
\newcommand{\GeV}{\,\mathrm{GeV}}
\newcommand{\MeV}{\,\mathrm{MeV}}
\newcommand{\amp}[1]{\mathcal{A}\left({#1}\right)}
\newcommand{\wilson}[2][{}]{\mathcal{C}_{#2}^{\mathrm{#1}}}
\newcommand{\bra}[1]{\left\langle{#1}\right\vert}
\newcommand{\ket}[1]{\left\vert{#1}\right\rangle}

%%%%
%    Greek Letters
%

\let\a=\alpha      \let\b=\beta       \let\c=\chi        \let\d=\delta
\let\e=\varepsilon \let\f=\varphi     \let\g=\gamma      \let\h=\eta
\let\k=\kappa      \let\l=\lambda     \let\m=\mu
\let\o=\omega      \let\r=\varrho     \let\s=\sigma
\let\t=\tau        \let\th=\vartheta  \let\y=\upsilon    \let\x=\xi
\let\z=\zeta       \let\io=\iota      \let\vp=\varpi     \let\ro=\rho
\let\ph=\phi       \let\ep=\epsilon   \let\te=\theta
\let\n=\nu
\let\D=\Delta   \let\F=\Phi    \let\G=\Gamma  \let\L=\Lambda
\let\O=\Omega   \let\P=\Pi     \let\Ps=\Psi   \let\Si=\Sigma
\let\Th=\Theta  \let\X=\Xi     \let\Y=\Upsilon
%
%%%

%%%
%    Calligraphic letters

\def\cA{{\cal A}}                \def\cB{{\cal B}}
\def\cC{{\cal C}}                \def\cD{{\cal D}}
\def\cE{{\cal E}}                \def\cF{{\cal F}}
\def\cG{{\cal G}}                \def\cH{{\cal H}}
\def\cI{{\cal I}}                \def\cJ{{\cal J}}
\def\cK{{\cal K}}                \def\cL{{\cal L}}
\def\cM{{\cal M}}                \def\cN{{\cal N}}
\def\cO{{\cal O}}                \def\cP{{\cal P}}
\def\cQ{{\cal Q}}                \def\cR{{\cal R}}
\def\cS{{\cal S}}                \def\cT{{\cal T}}
\def\cU{{\cal U}}                \def\cV{{\cal V}}
\def\cW{{\cal W}}                \def\cX{{\cal X}}
\def\cY{{\cal Y}}                \def\cZ{{\cal Z}}

%%%%

\def\be{\begin{equation}}
\def\ee{\end{equation}}
\def\bea{\begin{eqnarray}}
\def\eea{\end{eqnarray}}
\def\bm{\begin{matrix}}
\def\em{\end{matrix}}
\def\bpm{\begin{pmatrix}}
    \def\epm{\end{pmatrix}}

{\newcommand{\lsim}{\mbox{\raisebox{-.6ex}{~$\stackrel{<}{\sim}$~}}}
{\newcommand{\gsim}{\mbox{\raisebox{-.6ex}{~$\stackrel{>}{\sim}$~}}}
\def\mpl{M_{\rm {Pl}}}
\def\gev{{\rm \,Ge\kern-0.125em V}}
\def\tev{{\rm \,Te\kern-0.125em V}}
\def\mev{{\rm \,Me\kern-0.125em V}}
\def\ev{\,{\rm eV}}

\title{\boldmath  \color{schrift}{$B\to K^\ast_2(1430)\ell^+\ell^-$ distributions in  Standard Model at large recoil}}
\author{Diganta Das}
\email{diganta99@gmail.com}
\affiliation{Department of Physics and Astrophysics, University of Delhi, Delhi 110007, India}
\author{Bharti Kindra}
\email{bharti@prl.res.in}
\affiliation{Physical Research Laboratory, Navrangpura, Ahmedabad 380 009, India}
\affiliation{Indian Institute of Technology Gandhinagar, Gandhinagar 382 424, India}

\author{Girish Kumar}
\email{girishk@theory.tifr.res.in}
\affiliation{Department of Theoretical Physics, Tata Institute of Fundamental Research, 1 Homi Bhabha Road, Mumbai 400005, India}
\author{Namit Mahajan}
\email{nmahajan@prl.res.in}
\affiliation{Physical Research Laboratory, Navrangpura, Ahmedabad 380 009, India}

\begin{abstract}
We study the  rare decay $B\to K_2^\ast(1430)(\to K\pi)\ell^+\ell^-$ in the Standard Model and beyond.  Working in the transversity basis, we exploit the relations between the heavy-to-light form factors in the limit of heavy quark ($m_b\to \infty$) and large energy ($E_{K_2^\ast}\to \infty$) of the $K^\ast_2$ meson. This allows us to construct observables where at leading order in $\Lambda_{\rm QCD}/m_b$ and $\alpha_s$ the form factor dependence involving the $B\to K^\ast_2$ transitions cancels. Higher order corrections are systematically incorporated in the numerical analysis. In the Standard Model the decay has a sizable branching ratio and therefore a large number of events can be expected at LHCb.  Going beyond the Standard Model, we explore the implications of the global fit to presently available $b\to s\ell^+\ell^-$ data on the $B\to K_2^\ast \ell^+\ell^-$ observables.

\end{abstract}

\maketitle
%%%%%%%%%%%%%%%%%%%%%%%%%%%%%%%%%%%%%%%%%%%%%%%%%%%%%%%%%%%%%%%%%%%%%%%%%%%%%%%%%%%%%%%%%%%%%%%%%%%%%%%%%%%%%%%%%%%%%%%%%%%%%%%%%%%%%%%%%%%%%%%%%%%%%%%%
\section{Introduction \label{sec:intro}}
In the ongoing endeavor to unravel the flavor structure at the electroweak scale, the $b$-flavored mesons have played a very important role. In this effort, the exclusive $B$ meson decays that are induced by the $b\to s\ell^+\ell^-$ flavor changing neutral current transition are sensitive to physics in and beyond the Standard Model (SM), also known as the New Physics (NP). Well known candidates of this type of decays, the $B\to (K, K^\ast) \ell^+ \ell^-$ are at the center of theoretical and experimental investigations at present. Recent measurements of the observables $R_{K^{(\ast)}} \equiv {\rm BR} (B\to K^{(\ast)} \mu^+\mu^-)/{\rm BR} (B\to K^{(\ast)} e^+e^-)$ have shown hints of violation of lepton flavor universality (LFU). More explicitly, the LHCb collaboration has measured $R_{K}^{\rm{LHCb}}=0.745\pm_{0.074}^{0.090}\pm 0.036$ in $q^{2}\in [1,6]~\text{GeV}^{2}$ \cite{Aaij:2014ora},  $R_{K^\ast}^{\rm{LHCb}} = 0.66 ^{+0.11}_{-0.07} \pm 0.03$  in $q^{2}\in [0.045, 1.1]~\text{GeV}^{2}$, and $R_{K^\ast}^{\rm{LHCb}} = 0.69 ^{+0.11}_{-0.07} \pm 0.05$ in $q^{2}\in [1.1, 6]~\text{GeV}^{2}$ \cite{Aaij:2017vbb} and finds departure from the SM prediction $R_{K^{(\ast)}}\sim 1$ by about $\sim 2-3 \sigma$. Other notable deviation is the anomaly in the $P_5^\prime$ \cite{Descotes-Genon:2013wba} observed in recent measurements \cite{Aaij:2013qta,Aaij:2015oid,Abdesselam:2016llu,Wehle:2016yoi,ATLAS:2017dlm,CMS:2017ivg}, and the systematic deficit in $B_s\to \phi \mu^+\mu^-$ branching ratio \cite{Aaij:2015esa}. Global fits to these $b\to s\ell^+\ell^-$ data \cite{Descotes-Genon:2013wba,Descotes-Genon:2015uva,Altmannshofer:2014rta,Hurth:2016fbr,Altmannshofer:2015sma,Altmannshofer:2013foa,Beaujean:2013soa,Hurth:2013ssa} suggest that NP contributions to the Wilson coefficients can alleviate some of these tensions.

If the anomalies are indeed due to NP, it will show up in other $b\to s\ell^+\ell^-$ mediated transitions as well. The decay $B\to K^\ast_2(1430)\ell^+\ell^-$, where $K_2^\ast$ is a tensor meson is very similar to the well studied $B\to K^\ast\ell^+\ell^-$ and can provide complementary information to NP. Interestingly, the closely related radiative mode $B\to K_2^\ast\gamma$ has already been observed by the Babar~\cite{Aubert:2003zs} and Belle \cite{Nishida:2002me} collaborations and the branching ratio is comparable to that with $B\to K^\ast\gamma$. This implies that the mode $B\to K_2^\ast\ell^+\ell^-$ also has sizable branching ratio which has been confirmed by direct computations \cite{RaiChoudhury:2006bnu,Choudhury:2009fz,Hatanaka:2009gb,Hatanaka:2009sj}. 

The short distance physics of $B\to K_2^\ast\ell^+\ell^-$ is contained in the perturbatively calculable Wilson coefficients. The long-distance physics of $B\to K_2^\ast$ hadronic matrix elements are parametrized in terms of the form factors and the parametrization is similar to that of $B\to K^\ast$ hadronic matrix elements \cite{Hatanaka:2009sj}. The form factors has been calculated \cite{Wang:2010ni} in perturbative QCD approach using light-cone distribution amplitudes \cite{Cheng:2010hn} and in light-cone sum rules in conjunction with the $B$ meson wave function \cite{Wang:2010tz}. Calculations in light-cone QCD sum rule approach can be found in \cite{Yang:2010qd}. Using different  form factors,   phenomenological analysis of $B\to K_2^\ast\ell^+\ell^-$ has been performed in many works \cite{Ahmed:2012zzc,Junaid:2011bh,Hatanaka:2009sj,Li:2010ra,Lu:2011jm,Aliev:2011gc,RaiChoudhury:2006bnu}.  Majority of these works  have focussed on simple observables like decay rate, forward-backward asymmetry of dilepton system,  and the  polarization fractions of $K_2^\ast$.  In Ref.~\cite{RaiChoudhury:2006bnu}, the four-fold angular distribution of decay products of $K_2^\ast$ has been analysed in the SM. In Ref.~\cite{Li:2010ra}, the decay $B\to K_2^\ast(\to K\pi)\ell^+\ell^-$   was studied in the SM as well as in non-universal $Z^\prime$ and vector-like quark models, but branching fraction of the decay $K_2^\ast\to K\pi$ was ignored in their analysis. In this paper, building upon the previous works  we study the full four fold angular distribution of  $B\to K_2^\ast (\to K\pi)\ell^+\ell^-$ decay in the low dilepton invariant mass squared $q^2$ or large recoil of the $K^\ast_2$ meson. In this region, the heavy quark ($m_b\to \infty$) and large recoil ($E_{K_2^\ast}\to \infty$)  imply relations between $B\to K^\ast_2$ form factors. These relations reduce the number of independent form factors from seven to two. This helps us construct ``clean" observables where the form factor dependence cancels at the leading order in $\Lambda_{\rm QCD}/m_b$ and $\alpha_s$, making them suitable probes of NP. We have presented the determinations of the clean observables in the SM and studied the implications of global fits to the present $b\to s\ell^+\ell^-$ data.

The rest of the paper is organized as follows. In Sec.~\ref{subsec:effHam} we discuss the general effective Hamiltonian and relevant operators for $b \to s \ell^+\ell^-$. In the Sec.~\ref{subsec:fullFF} the hadronic matrix elements for $B \to K_2^\ast $ and their parametrization in terms of form factors are discussed. In Sec.~\ref{subsec:TAs} we discuss the $B \to K_2^\ast $ helicity amplitudes in the transversity basis and give their expressions in terms of form factors and short-distance Wilson coefficients. In Sec.~\ref{subsec:largeRec} we discuss  $B\to K_2^\ast \ell^+\ell^-$ in the large recoil and large energy limit in detail. In Sec.~\ref{sec:angular} the four-body fully differential angular distribution and angular observables for $B\to K_2^\ast (\to K\pi) \ell^+\ell^-$ are discussed.  In the next Sec.~\ref{sec:glfit} we discuss the considered angular observables in the SM and in interesting NP scenarios. We give numerical predictions for observables and discuss their sensitivity to possible NP in $b\to s \ell^+\ell^-$. Finally in Sec.~\ref{Sec:summary} we summarize our results of this paper.

%%%%%%%%%%%%%%%%%%%%%%%%%%%%%%%%%%%%%%%%%%%%%%%%%%%%%%%%%%%%%%%%%%%%%%%%%%%%%%%%%%%%%%%%%%%%%%%%%%%%%%%%%%%%%%%%%%%%%%%%%%%%%%%%%%%%%%%%%%%%%%%%%%%%%%%%
\section{Effective Hamiltonian \label{subsec:effHam}}
We work with the following low energy effective Hamiltonian for rare $|\Delta B|=|\Delta S|=1$ transition 
\begin{eqnarray}
  \label{eq:Heff}
  {\cal{H}}_{\rm eff}=
   - \frac{4\, G_F}{\sqrt{2}}  V_{tb}^{} V_{ts}^\ast \,\frac{\alpha_e}{4 \pi}\,
     \sum_{i=7,9,10} \Big[ C_i(\mu)  {\cal{O}}_i(\mu) + C_i^\prime(\mu)  {\cal{O}}_i^\prime(\mu)\Big]\, ,
\end{eqnarray}
where
\begin{eqnarray}
 {\cal{O}}_{7}^{(\prime)} &=& \frac{m_b}{e} [\bar{s} \sigma^{\mu\nu} P_{R}(P_L) b]  F_{\mu\nu}\,, \quad
 {\cal{O}}_{9}^{(\prime)} =  [\bar{s} \gamma_\mu P_{L}(P_R) b] \, [\bar{\ell} \gamma^\mu \ell] \,, \quad
 {\cal{O}}_{10}^{(\prime)} = [\bar{s} \gamma_\mu P_{L}(P_R) b] \, [\bar{\ell} \gamma^\mu \gamma_5 \ell] \,.
\end{eqnarray}
%%%%%
Here $\mu$ is the renormalization scale, $\alpha_e$ is the fine structure constant, $F_{\mu\nu}$ is the electromagnetic field
strength tensor and $P_{L/R}=(1 \mp \gamma_5)/2$ are the chiral projectors. The $b$-quark mass multiplying the dipole operator is assumed to be the running quark mass in the modified minimal-subtraction ($\overline{\rm MS}$) mass scheme. The contributions of the factorizable quark-loop corrections to current-current and penguin operators are absorbed in the effective Wilson coefficients $C^{\rm eff}_{7,9}$ as described in Appendix~\ref{app:C79eff}. All the SM Wilson coefficients are evaluated at the renormalization scale $\mu = m_b = 4.2$ GeV \cite{Detmold:2016pkz}. For simplicity we will neglect the superscript ``eff" in the rest of the text. We ignore the non-factorizable corrections to the Hamiltonian which are expected to be significant at large recoil \cite{Beneke:2001at,Beneke:2004dp}. The primed Wilson coefficients are zero in the SM but can appear in some NP models. We will not consider NP contributions to $\mathcal{O}_7$ since these are well constrained \cite{Paul:2016urs}. 
%%%%%%%%%%%%%%%%%%%%%%%%%%%%%%%%%%%%%%%%%%%%%%%%%%%%%%%%%%%%%%%%%%%%%%%%%%%%%%%%%%%%%%%%%%%%%%%%%%%%%%%%%%%%%%%%%%%%%%%%%%%%%%%%%%%%%%%%%%%%%%%%%%%%%%%%%%%%%%%%%%%%%%%%%%%%%%%%%%%%%%
\section{$B\to K_2^\ast$ hadronic matrix elements \label{subsec:fullFF}}
We work in the $B$ meson rest frame and denote by $p,k,p_{\ell^+}, p_{\ell^-}$ the four-momentum of the $B$-meson, the $K^\ast_2$, and the positively and negatively charged leptons respectively. Tensor meson of spin-2 polarization tensor $\epsilon^{\mu\nu}(n)$, where the helicities are $n=t, 0,\pm 1, \pm 2$, satisfies $\epsilon^{\mu\nu}k_\nu=0$. In the final state, the $K^\ast_2$ meson is partnered with two spin-half leptons and hence the $K^\ast_2$ can only have helicities $n=t, 0,\pm 1$.  Noting that the polarization tensor of spin-$2$ state $K_2^\ast$ can be conveniently written in terms of polarization vectors of a spin-$1$ state \cite{Berger:2000wt}, we  introduce a new polarization vector $\epsilon_{T\mu}$ (see Appendix.~\ref{app:polz}) in terms of which the $B\to K^\ast_2$ hadronic matrix elements can be written as \cite{Wang:2010ni}
 \begin{eqnarray}\label{eq:B2K2ff}
  \langle K_2^*(k,n)|\bar s\gamma^{\mu}b|\overline B(p)\rangle
  &=&-\frac{2V(q^2)}{m_B+m_{K_2^*}}\epsilon^{\mu\nu\rho\sigma} \epsilon^*_{T\nu}  p_{\rho}k_{\sigma}, \nonumber\\
  \langle  K_2^*(k,n)|\bar s\gamma^{\mu}\gamma_5 b|\overline B(p)\rangle
   &=&2im_{K_2^*} A_0(q^2)\frac{\epsilon^*_{T } \cdot  q }{ q^2}q^{\mu} + i(m_B+m_{K_2^*})A_1(q^2)\left[ \epsilon^*_{T\mu }
    -\frac{\epsilon^*_{T } \cdot  q }{q^2}q^{\mu} \right] \nonumber\\
    &&-iA_2(q^2)\frac{\epsilon^*_{T} \cdot  q }{  m_B+m_{K_2^*} }
     \left[ P^{\mu}-\frac{m_B^2-m_{K_2^*}^2}{q^2}q^{\mu} \right],\\
  \langle  K_2^*(k,n)|\bar sq_{\nu}\sigma^{\mu\nu}b|\overline B(p)\rangle
   &=&-2iT_1(q^2)\epsilon^{\mu\nu\rho\sigma} \epsilon^*_{T\nu} p_{B\rho}p_{K\sigma}, \nonumber\\
  \langle  K_2^*(k,n)|\bar sq_{\nu}\sigma^{\mu\nu}\gamma_5b|\overline  B(p)\rangle
   &=&T_2(q^2)\left[(m_B^2-m_{K_2^*}^2) \epsilon^*_{T\mu }
       - {\epsilon^*_{T } \cdot  q }  P^{\mu} \right] +T_3(q^2) {\epsilon^*_{T } \cdot  q }\left[
       q^{\mu}-\frac{q^2 (p+k)^{\mu}}{m_B^2-m_{K_2^*}^2}\right]\, ,\nn 
 \end{eqnarray}
where $q = p_{\ell^+} + p_{\ell^-} = p-k$ is the momentum transfer.

%%%%%%%%%%%%%%%%%%%%%%%%%%%%%%%%%%%%%%%%%%%%%%%%%%%%%%%%%%%%%%%%%%%%%%%%%%%%
%%%%%%%%%%%%%%%%%%%%%%%%%%%%%%%%%%%%%%%%%%%%%%%%%%%%%%%%%%%%%%%%%%%%%%%%%%%%
\section{Transversity amplitudes \label{subsec:TAs}}
Corresponding to the effective Hamiltonian (\ref{eq:Heff}), the $B\to K_2^\ast\ell^+\ell^-$ amplitude for a given helicity of the $K_2^\ast$ can be written as
\begin{eqnarray}
\mathcal{A}(n) &=& -\frac{G_F}{\sqrt{2}}V_{tb}V_{ts}^\ast \frac{\alpha_e}{\pi}
\Bigg( \Bigg [(C_9 - C_{10}) \langle K_2^*(k,n)|\bar s\gamma^{\mu} P_L b|\overline B(p)\rangle -
i\frac{2C_7 m_b}{q^2} \langle K_2^*(k,n)|\bar s \sigma^{\mu\nu}q_\nu  P_R  b|\overline B(p)\rangle\, \nn\\
&+& (C_9^\prime - C_{10}^\prime) \langle K_2^*(k,n)|\bar s\gamma^{\mu} P_R b|\overline B(p)\rangle \Bigg] \bar{\ell}\gamma_\mu P_L \ell + \bigg[C_{10}\to -C_{10}, C^\prime_{10}\to -C_{10}^\prime \bigg] \bar{\ell}\gamma_\mu P_R \ell \Bigg) \, .
\end{eqnarray}
The differential distribution for the decay can be calculated using helicity amplitudes $H^{L,R}_\pm$ and $H^{L,R}_0$ which are defined as the projections of the hadronic amplitudes on the polarization vectors of the gauge boson that creates dilepton pair. Here the superscripts $L,R$ correspond to the chiralities of the leptonic current. However, for comparison with the literature we introduce the so called the transversity amplitudes which are linear combinations of helicity amplitudes: $A_{\|L,R} = (H^{L,R}_+ + H^{L,R}_-)/\sqrt{2}$,
$A_{\perp L,R} = (H^{L,R}_+ - H^{L,R}_-)/\sqrt{2}$, and $A_{0L,R} = H^{L,R}_0$. The expressions of the transversity amplitudes for $B\to K_2^\ast(\to K\pi)\ell^+\ell^-$ read \cite{Li:2010ra}
\begin{eqnarray}\label{eq:TAFull}
 A_{0L,R}
  &=& N  \frac{\sqrt{\lambda}}{\sqrt6 m_Bm_{K_2^*}}\frac{1}{2m_{K^*_2}\sqrt {q^2}}\left[ (C_{9-}\mp C_{10-})
[(m_B^2-m_{K^*_2}^2-q^2)(m_B+m_{K^*_2})A_1
 -\frac{\lambda}{m_B+m_{K^*_2}}A_2]\right.\nonumber\\
 &&\left. +  2m_b C_7  [ (m_B^2+3m_{K_2^*}^2-q^2)T_2 -\frac{\lambda  }
 {m_B^2-m_{K_2^*}^2}T_3]\right],\\
 A_{\perp L,R}&=& -\sqrt{2} \frac{\sqrt{\lambda}}{\sqrt8m_Bm_{K_2^*}}N\left[(C_{9+}\mp C_{10+})
 \frac{\sqrt \lambda V}{m_B+m_{K^*_2}}+\frac{2m_b C_{7}}{q^2}\sqrt \lambda T_1\right],\\
 A_{|| L,R}&=& \sqrt{2}\frac{\sqrt{\lambda}}{\sqrt 8m_Bm_{K_2^*}} N \left[(C_{9-}\mp C_{10-}) (m_B+m_{K^*_2})A_1+\frac{2m_b C_{7}}{q^2}(m_B^2-m_{K^*_2}^2)
 T_2 \right]\, ,\\
 A_t &=& 2\frac{\sqrt{\lambda}}{\sqrt{6}m_B m_{K^\ast_2}} N [C_{10-}] A_0 \, ,
\end{eqnarray}
where the normalization constant is given by
\begin{equation}
 N = \Bigg[ \frac{G_F^2\alpha_e^2}{3\cdot 2^{10}\pi^5m_B^3}|V_{tb}V_{ts}^\ast|^2 \lambda^{1/2}(m_B^2, m_{K_2^\ast}^2, q^2) \mathcal{B}(K_2^\ast\to K\pi)\beta_\ell \Bigg]^\frac{1}{2}\, ,\quad \beta_\ell = \sqrt{1 - \frac{4m_\ell^2}{q^2}}\, .
\end{equation}
Here $\lambda(a,b,c) = a^2 + b^2 + c^2 -2(ab + bc + ca)$ and we have defined
\begin{eqnarray}
C_{9\pm} = C_{9}  \pm C_{9}^\prime\, ,\quad C_{10\pm} = C_{10} \pm C_{10}^\prime\, .
\end{eqnarray}

\section{Heavy to light form factors at large recoil \label{subsec:largeRec}}
The $B\to K_2^\ast$ hadronic matrix elements, parametrized in terms of form factors in eq.~(\ref{eq:B2K2ff}) are non-perturbative in nature and constitute the dominant uncertainty in theoretical predictions. In the absence of any lattice calculations of the form factors at present, the uncertainty can be reduced by making use of the relations between the form factors that originate in the limit of heavy quark $m_b\to\infty$ of the initial meson and large energy $E_{K_2^\ast}$ of the final meson \cite{Dugan:1990de,Charles:1998dr}. In these limits, the heavy to light form factors can be expanded in small ratios of $\Lambda_{\rm QCD}/m_b$ and $\Lambda_{\rm QCD}/E_{K^\ast_2}$. To leading order in $\Lambda_{\rm QCD}/m_b$ and $\alpha_s$, the large energy symmetry dictates that there are only two independent universal soft form factors, $\xi_\|(q^2)$ and $\x_\perp(q^2)$ \cite{Charles:1998dr}, in terms of which the rest of the form factors can be written as \cite{Hatanaka:2009sj} 
\begin{equation}\label{eq:softFull}
\begin{split}
& A_0(q^2) = \frac{\mK}{|p_{K^\ast_2}|} \Bigg[ \Bigg( 1-\frac{\mKK}{m_B} \Bigg)\xi_\|(q^2) + \frac{\mK}{m_B}\xi_\perp(q^2) \Bigg]\, ,\\
& A_1(q^2) = \frac{\mK}{|p_{K^\ast_2}|} \frac{2\EK}{m_B+\mK}\xi_\perp(q^2)\, , \quad
A_2(q^2) = \frac{\mK}{|p_{K^\ast_2}|} \Big(1 + \frac{\mK}{m_B} \Big) \Big[\xi_\perp(q^2) - \frac{\mK}{E} \xi_\| \Big]\, ,\\
&V(q^2) = \frac{\mK}{|p_{K^\ast_2}|} \Big(1 + \frac{\mK}{m_B} \Big) \xi_\perp\, ,\quad
T_1(q^2) = \frac{\mK}{|p_{K^\ast_2}|} \xi_\perp(q^2)\, ,\\
&T_2(q^2) = \frac{\mK}{|p_{K^\ast_2}|} \Big(1 - \frac{q^2}{m_B^2-\mKK} \Big) \xi_\perp(q^2)\, ,\quad
T_3(q^2) = \frac{\mK}{|p_{K^\ast_2}|} \Big[ \xi_\perp - \Big(1 - \frac{\mKK}{m_B^2} \Big) \frac{\mK}{E} \xi_\|(q^2) \Big] \, .
\end{split}
\end{equation}
Here recoil energy $\EK$ is given by
\begin{equation}
\EK = \frac{m_B}{2}\Bigg( 1 - \frac{q^2}{m_B^2} + \frac{\mKK}{m_B^2} \Bigg)\, .
\end{equation}
The $q^2$ dependence of the soft form factors $\xi_{\perp}(q^2)$ and $\xi_{\|}(q^2)$   is given by \cite{Charles:1998dr,Hatanaka:2009sj}
\begin{equation}\label{eq:xi}
\xi_{\|,\perp}(q^2) = \frac{\xi_{\|,\perp}(0)}{(1-q^2/m_B^2)^2}\,.
\end{equation}
The values of the soft form factors at the zero recoil $q^2=0$ have been estimated using Bauer-Stech-Wirbel (BSW) model \cite{Wirbel:1985ji} in Ref.~\cite{Hatanaka:2009sj}. In Ref.~\cite{Hatanaka:2009gb} these are also extracted from experimental data on $B\to K^\ast\gamma$ from Babar~\cite{Aubert:2003zs} and Belle \cite{Nishida:2002me}. For our numerical analysis we have used the values $\xi_\perp(0)=0.29\pm 0.09$ and $\xi_\|(0)=0.26\pm 0.10$ which were obtained in Ref.~\cite{Wang:2010ni} in perturbative QCD approach utilizing the non-trivial relations realized in the large energy limit. These estimates are consistent with the ones obtained in Refs.~\cite{Hatanaka:2009sj} and \cite{Hatanaka:2009gb} but have higher errors. To be more conservative in our theory estimates, we choose to use values given above.

Substituting (\ref{eq:softFull}) in (\ref{eq:TAFull}), we obtain at leading order in $\Lambda_{\rm QCD}/m_b$ and $\alpha_s$ the simple expressions of the transversity amplitudes in terms of soft form factors $\xi_\|$ and $\xi_\perp$ as
\begin{eqnarray}
A_{0L(R)} &=& \sqrt{\frac{2}{3}} \frac{N}{\sqrt{q^2}} m_B^2 \Big(1-\frac{q^2}{m_B^2}\Big) \Bigg( (C_{9-} \mp C_{10+}) + 2C_{7}  \frac{m_b}{m_B} \Bigg) \xi_\|(q^2)\, ,\label{eq:TAsinSFF1}\\
A_{\perp L(R)} &=& -N m_B \Big(1-\frac{q^2}{m_B^2}\Big) \Bigg( (C_{9+}\mp C_{10+}) + 2C_{7} \frac{m_b m_B}{q^2} \Bigg) \xi_\perp(q^2)\, ,\label{eq:TAsinSFF2}\\
A_{\| L(R)} &=& N m_B \Big(1-\frac{q^2}{m_B^2}\Big) \Bigg( (C_{9-}\mp C_{10-}) + 2C_{7} \frac{m_b m_B}{q^2} \Bigg) \xi_\perp(q^2)\, ,\label{eq:TAsinSFF3}\\
 A_t &=& 2\frac{\sqrt{\lambda}}{\sqrt{6}m_B m_{K^\ast_2}} N  \frac{2 m_{K_2^\ast} m_B}{\sqrt{\lambda}} \Bigg( (1 - \frac{m_{K^\ast_2}^2}{m_B E_{K_2^\ast} })\xi_\| + \frac{m_{K^\ast_2}}{m_B} \xi_\perp \Bigg) C_{10-} \,\label{eq:TAsinSFF4}.
\end{eqnarray}

At this point we recall that the relations (\ref{eq:softFull}) are derived in the QCD factorization (QCDf) and soft-collinear-effective theory (SCET) approach in which the factorization formula for the heavy to light $B\to K^\ast_2$ form factors are
\begin{equation}
	F_i(q^2) = (1+\mathcal{O}(\alpha_s))\xi + \Phi_B \oplus T_i \oplus \Phi_{K_2^\ast} + \mathcal{O}(\Lambda_{\rm QCD}/m_b)\, .
\end{equation} 
Here $T_i$ are the perturbatively calculable hard scattering kernals and $\Phi_{B,K^\ast}$ are the hadron distribution amplitudes which are non perturbative objects. There are no means to calculate the $\Lambda_{\rm QCD}/m_b$ corrections at present, and therefore the cancellations of soft form factors in the clean observables are valid only at leading order in $\Lambda_{\rm QCD}/m_b$. The neglected higher order terms add to the uncertainty of our theoretical predictions. We use the ensemble method following  Ref.~\cite{Egede:2010zc} to account for  $\Lambda_{\rm QCD}/m_b$  uncertainties in our numerical analysis  of observables. This is done by multiplying the transversity amplitudes by correction factors
\begin{equation}\label{eq: lQCD correction}
	A_{0,\|,\perp} \to A_{0,\|,\perp}(1+c_{0,\|,\perp})\, ,
\end{equation} 
where $c_{0,\|,\perp}$ are the correction factors defined as $c_{0,\|,\perp} = |c_{0,\|,\perp}|e^{i\theta_{0,\|,\perp}}$. We vary $|c_{0,\|,\perp}|$ and $\theta_{0,\|,\perp}$  in a  random uniform distribution in the ranges $[-0.1,0.1]$ and $[-\pi,\pi]$  respectively.  Other sources of uncertainties are the due to the  variation of scale $\mu$ between $m_b/2$ - $2m_b$ and the ratio $m_c/m_b$. Some of the inputs and their uncertainties are listed in Table \ref{tab:input}.

\section{Angular distributions and observables \label{sec:angular}}
We assume that the $K_2^\ast$ is on the mass shell so that the $B\to K_2^\ast(\to K\pi)\ell^+\ell^-$ decay can be completely described in terms of only four kinematical variables; the lepton invariant mass squared $q^2$ and three angles $\theta_\ell, \theta_K$ and $\phi$. The lepton angle $\theta_\ell$ is defined as the angle made by the negatively charged lepton $\ell^-$ with respect to the direction of the motion of the $B$ meson in the di-lepton rest frame. The angle $\theta_K$ is defined as the angle made by the $K^-$ with respect to the opposite of the direction of the $B$ meson in the $K\pi$ rest frame. The angle between the decay planes of the two leptons and the $K\pi$ is defined as $\phi$. In terms of these variables, the fourfold differential distributions read \cite{Li:2010ra}
\begin{eqnarray}\label{eq:diff4}
\frac{d^4\Gamma}{dq^2 d\cos\theta_\ell d\cos\theta_K d\phi} &=& \frac{15}{128\pi}\Bigg[ I_1^s 3\sin^2 2\theta_K + I_1^c (3\cos^2 \theta_K-1)^2 +  I_2^s 3\sin^22\theta_K \cos 2\theta_l \,\nn\\ &+& I_2^c (3\cos^2\theta_K-1)^2 \cos 2\theta_l + I_3  3\sin^22\theta_K \sin^2\theta_l \cos 2\phi \,\nn\\ &+& I_4 2\sqrt{3} (3\cos^2\theta_K-1)\sin 2\theta_K \sin 2\theta_l \cos\phi \,\nn\\ &+& I_5 2\sqrt{3} (3\cos^2\theta_K-1)\sin 2\theta_K \sin \theta_l \cos\phi + I_6 3 \sin^22\theta_K \cos\theta_l \,\nn\\ &+& I_7 2\sqrt{3} (3\cos^2\theta_K-1)\sin 2\theta_K \sin \theta_l \sin\phi\, \nn\\ &+& I_8 2\sqrt{3} (3\cos^2\theta_K-1)\sin 2\theta_K \sin 2\theta_l \sin\phi \,\nn\\ &+& I_9 3 \sin^22\theta_K \sin^2\theta_l \sin 2\phi \Bigg]
\end{eqnarray}
The angular coefficients $I_i(q^2)$ can be written in terms of the transversity amplitudes and are given in Appendix~\ref{app:Is}. The decay rate for the CP-conjugate process is obtained by the replacements $I_{1,2,3,4,7} \to \bar{I}_{1,2,3,4,7}$ and $I_{5,6,8,9}\to -\bar{I}_{5,6,8,9}$, where $\bar{I}$ are equal to $I$ with all the weak phase conjugated. In this paper we will consider only CP-averaged observables, so that $I$ means $I+\bar{I}$ and total decay width $\Gamma$ stands for $\Gamma+\bar{\Gamma}$.  At leading order $\Lambda_{\rm QCD}/m_b$ and $\alpha_s$ the short- and long-distance physics factorize as %
\begin{eqnarray}\label{eq:low-recoil-Is}
I_1^c&=&  \frac{2}{3}\frac{N^2}{q^2} m_B^4\left(1-\frac{q^2}{m_B^2}\right)^2\left[|\sigma_-|^2 + |\sigma_+|^2 + \frac{8\, m_\ell^2}{q^2}\{{\rm Re}(\sigma_-\sigma_+^\ast) +2|C_{10-}|^2\left( 1-\frac{2\,m_{K_2^\ast}^2}{m_B^2-q^2}+\frac{m_{K_2^\ast}}{m_B}\frac{\xi_\perp}{\xi_\parallel}\right)^2\}\right]\xi_\parallel^2,\nonumber\\
I_1^s&=&\frac{3 }{4}N^2 m_B^2\left(1-\frac{q^2}{m_B^2}\right)^2\left[\left(1-\frac{4\, m_\ell^2}{3 q^2}\right)\{(|\rho_-^L|^2+|\rho_+^L|^2) +(L\leftrightarrow R)\}+\frac{m_\ell^2}{3 q^2}\,{\rm Re}(\rho_-^L\rho_-^{R\ast}+\rho_+^L\rho_+^{R\ast})\right]\xi_\perp^2,\nonumber\\
I_2^c  &=&-\frac{2}{3}\frac{ N^2 }{ q^2}m_B^4\,\beta_\ell^2\left(1-\frac{q^2}{m_B^2}\right)^2(|\sigma_-|^2 + |\sigma_+|^2)\,\xi_\parallel^2,\nonumber\\
I_2^s  &=&\frac{1}{4}N^2m_B^2\,\beta_\ell^2\left(1-\frac{q^2}{m_B^2}\right)^2\{(|\rho_-^L|^2+|\rho_+^L|^2) +(L\leftrightarrow R)\}\xi_\perp^2\,,\nonumber\\
I_3  &=&\frac{1}{2}N^2m_B^2\,\beta_\ell^2\left(1-\frac{q^2}{m_B^2}\right)^2\{(|\rho_+^L|^2-|\rho_-^L|^2) +(L\leftrightarrow R)\}\xi_\perp^2\,,\nonumber\\
I_4
&=& \frac{1}{\sqrt{3}}\frac{N^2}{\sqrt{q^2}}m_B^3\,\beta_\ell^2\left(1-\frac{q^2}{m_B^2}\right)^2{\rm Re}[\sigma_-\rho_-^{L\ast}+\sigma_+\rho_-^{R\ast}]\xi_\perp\xi_\parallel\, ,\\
I_5
&=& -\frac{2}{\sqrt{3}}\frac{N^2}{\sqrt{q^2}}m_B^3\,\beta_\ell\left(1-\frac{q^2}{m_B^2}\right)^2{\rm Re}[\sigma_-\rho_+^{L\ast}-\sigma_+\rho_+^{R\ast}]\,\xi_\perp\xi_\parallel\, ,\nonumber\\
I_6
&=& -2N^2m_B^2\,\beta_\ell\left(1-\frac{q^2}{m_B^2}\right)^2{\rm Re}[(\rho_-^L\rho_+^{L\ast})-(L\leftrightarrow R)]\,\xi_\perp^2\,,\nonumber\\
I_7  &=& \frac{2}{\sqrt{3}}\frac{N^2}{\sqrt{q^2}}m_B^3\,\beta_\ell\left(1-\frac{q^2}{m_B^2}\right)^2{\rm Im}[\sigma_-\rho_-^{L\ast}-\sigma_+\rho_-^{R\ast}]\,\xi_\perp\xi_\parallel\, ,\nonumber\\
I_8 &=& -\frac{1}{\sqrt{3}}\frac{N^2}{\sqrt{q^2}}m_B^3\,\beta_\ell^2\left(1-\frac{q^2}{m_B^2}\right)^2{\rm Im}[\sigma_-\rho_+^{L\ast}+\sigma_+\rho_+^{R\ast}]\,\xi_\perp\xi_\parallel\, ,\nonumber\\
I_9
&=&-N^2m_B^2\,\beta_\ell^2\left(1-\frac{q^2}{m_B^2}\right)^2{\rm Im}[(\rho_-^L\rho_+^{L\ast})+(L\leftrightarrow R)]\,\xi_\perp^2\,.\nonumber
\end{eqnarray}
Here we have introduced the following combinations of short-distance Wilson coefficients
\begin{eqnarray}
	\rho_{\mp}^L(q^2) &=& C_{9\mp}-C_{10\mp}+\frac{2\,m_bm_B}{q^2}C_7,\\
	\rho_{\mp}^R(q^2) &=& C_{9\mp}+C_{10\mp}+\frac{2\,m_bm_B}{q^2}C_7,\\
	\sigma_\mp(q^2) &=& C_{9-} \mp C_{10+}+\frac{2\,m_b}{m_B}C_7.
\end{eqnarray}
While writing  Eq.~\eqref{eq:low-recoil-Is},  we have not displayed explicit $q^2$-dependence of form factors $\xi_{\perp,\parallel}(q^2)$ and the Wilson coefficients $\rho_{\mp}^{L,R}(q^2)$ for simplicity. Note that in the SM basis, one has $\rho_-^L=\rho_+^L$ and $\rho_-^R=\rho_+^R$.

From the angular distribution (\ref{eq:diff4})one can construct observables like the forward-backward asymmetry $A_{\rm FB}$, the longitudinal polarization fraction $F_L$, and the differential decay width $d\Gamma/dq^2$ as functions of dilepton invariant mass $q^2$. This can be done by weighted angular integrals of the four fold differential distribution given in Eq.~\eqref{eq:diff4} as the following
\begin{eqnarray}\label{eq:weighted}
{\cal O}_i(q^2) = \int d\cos\theta_\ell~d\cos\theta_K~d\phi~{\cal W}_i(\theta_\ell,\theta_K,\phi)\,\frac{d^4\Gamma}{dq^2 d\cos\theta_\ell d\cos\theta_K d\phi}\,,
\end{eqnarray}
from which various angular observables can be extracted by the suitable choices for weight function~${\cal W}_i(\theta_\ell,\theta_K,\phi)$:

\begin{itemize}
	\item  The full differential decay width $d\Gamma/dq^2$  is simply obtained by choosing ${\cal W}_\Gamma=1$,
	\begin{eqnarray}\label{eq:diffdecaywidth}
	%A_{\rm FB}(q^2) &=& \frac{3I_6}{3I_1^c + 6I_1^s - I_2^c - 2I_2^s} \, ,\\
	%F_L(q^2) &=& \frac{3I_1^c - I_2^c}{3I_1^c + 6I_1^s - I_2^c -2I_2^s}\, ,\\\label{eq:nc2}
	\frac{d\Gamma}{dq^2} &=& \frac{1}{4}(3I_1^c + 6I_1^s - I_2^c - 2I_2^s)\, .
	\end{eqnarray}
	\item The forward-backward asymmetry of lepton pair $A_{\rm FB}$ (normalized by differential decay width) is extracted with  ${\cal W}_{A_{\rm FB}}= {\rm sgn}[\cos\theta_\ell]/(d\Gamma/{dq^2})$,
	\begin{eqnarray}\label{eq:AFB}
	A_{\rm FB}(q^2) &=& \frac{3I_6}{3I_1^c + 6I_1^s - I_2^c - 2I_2^s}\,.
	\end{eqnarray}
	\item The longitudinal polarization fraction $F_L$ (normalized by differential decay width) is extracted with~${\cal W}_{F_L}=(3/2) (-3+7 \cos^2\theta_K)/(d\Gamma/{dq^2})$,
	\begin{eqnarray}\label{eq:FL}
	F_L(q^2) &=& \frac{3I_1^c - I_2^c}{3I_1^c + 6I_1^s - I_2^c -2I_2^s}\,.
	\end{eqnarray}
	By definition then the transverse polarization fraction is  $F_T=1-F_L$. 
\end{itemize}

In Table \ref{Table:normal_obs} we present our $q^2$-bin averaged estimates for the above observables for $B\to K_2^\ast \mu^+\mu^-$ in different bins in the SM. The sources of uncertainties are $\Lambda_{\rm QCD}/m_b$ corrections, variation of renormalization scale $\mu$, form factors and other numerical inputs. In Fig.~\ref{SM_obs}, we have shown the dependence of these three observables on dilepton invariant mass $q^2$. As can be seen from these analysis, the $B\to K^\ast_2\mu^+\mu^-$ branching ratio is only one order of magnitude smaller than $B\to K^\ast\mu^+\mu^-$. Therefore, $B\to K^\ast_2\mu^+\mu^-$ can be a viable signal at future LHCb. However, due to large uncertainties branching ratio, $A_{\rm FB}$ and $F_L$ are not suitable for searches of new physics.

\begin{table}[]
	\setlength{\tabcolsep}{5pt}
	\begin{tabular}{  c | c | c | c | c | c }
		\hline
		Observable \slash $q^2$ bin (${\rm GeV}^2$) & $[0.1 - 1.0]$ &$[1.0 - 2.0]$ &$[2.0 - 4.0]$ &$[4.0 - 6.0]$ &$[1.0 - 6.0]$ \\
		\hline
		$10^{7}\times\langle {\rm BR}(B \to K_2^\ast \mu\mu)\rangle$ & $0.205\pm 0.093$ &  $0.104 \pm 0.055$ &  $0.196\pm 0.119$&  $0.232 \pm 0.122$&  $0.532 \pm 0.289$ \\ 
		$\langle F_L \rangle$ & $0.346 \pm  0.198 $ &  $0.686 \pm 0.208$&  $0.760 \pm 0.191$&  $0.679 \pm 0.210$&  $0.709 \pm 0.204$   \\ 
		$\langle A_{\rm FB}\rangle$ &  $0.094 \pm 0.029$& $0.202 \pm 0.133$ &  $0.070 \pm .059$& $-0.141 \pm 0.094$&  $0.0 \pm 0.019$ \\
		\hline
	\end{tabular}
	\caption{Binned predictions for ${\rm BR}(B \to K_2^\ast \mu^+\mu^-)$ , $F_L$, and  $A_{FB}$ in the SM.  Theoretical errors correspond to uncertainties in the form factors,   $\Lambda_{\rm QCD}/m_b$ correction effects,  and errors in  inputs as discussed in the  text. }  
	\label{Table:normal_obs}
\end{table}
\begin{figure}[h!]
	\begin{center}\begin{tabular}{cc}
		\includegraphics[scale=0.4]{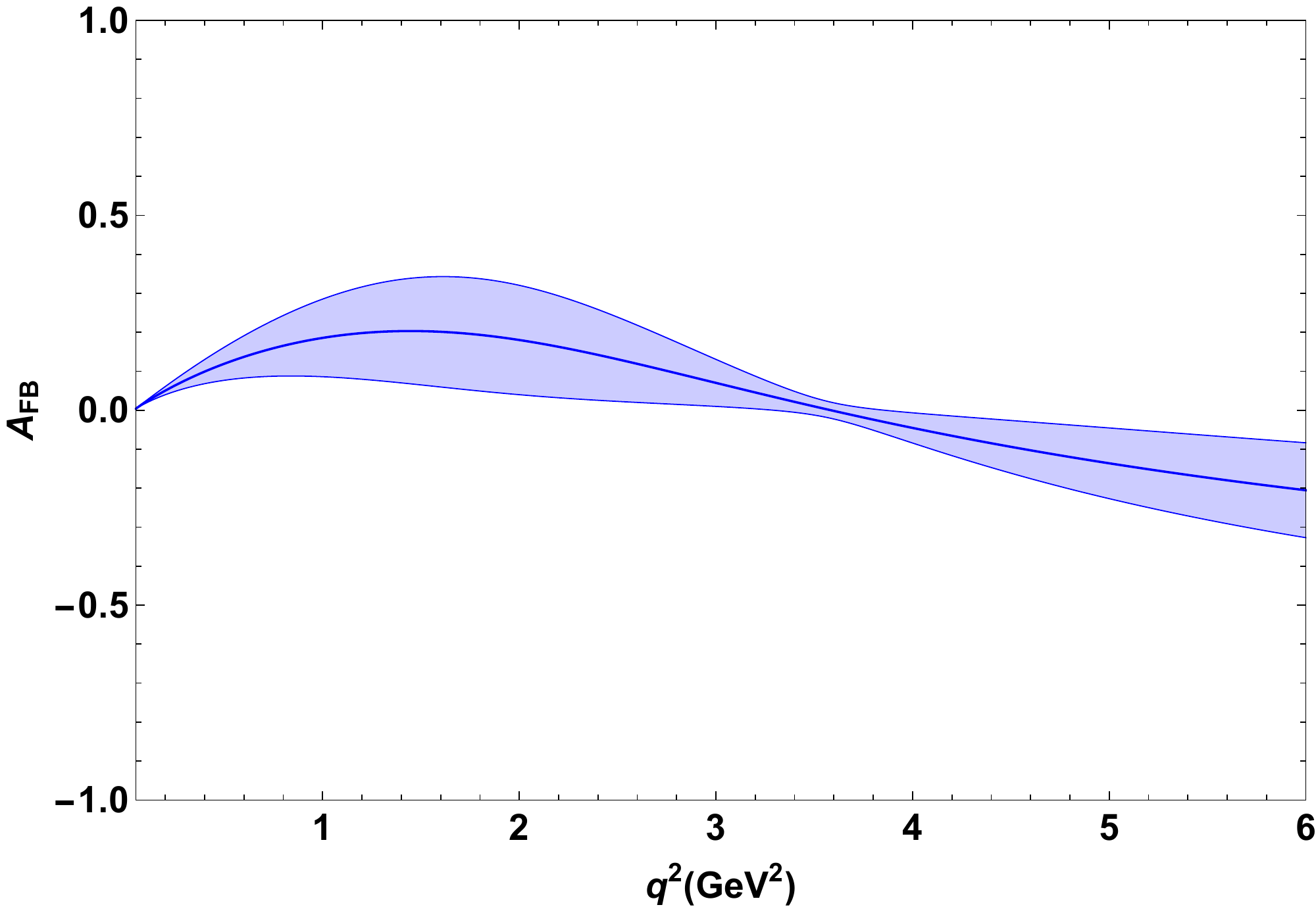}&
		\includegraphics[scale=0.4]{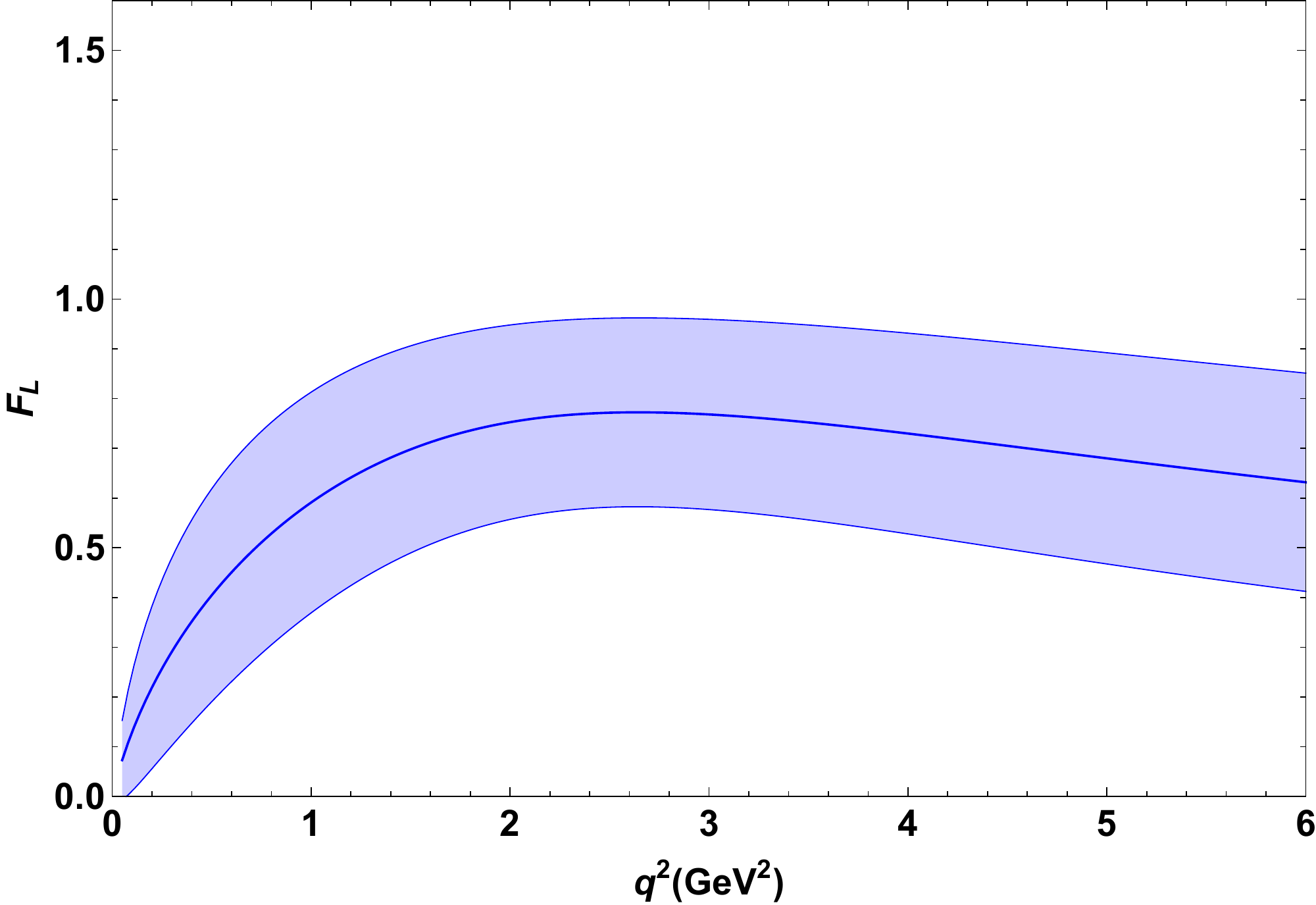}\\
			\end{tabular}\\
		\includegraphics[scale=0.4]{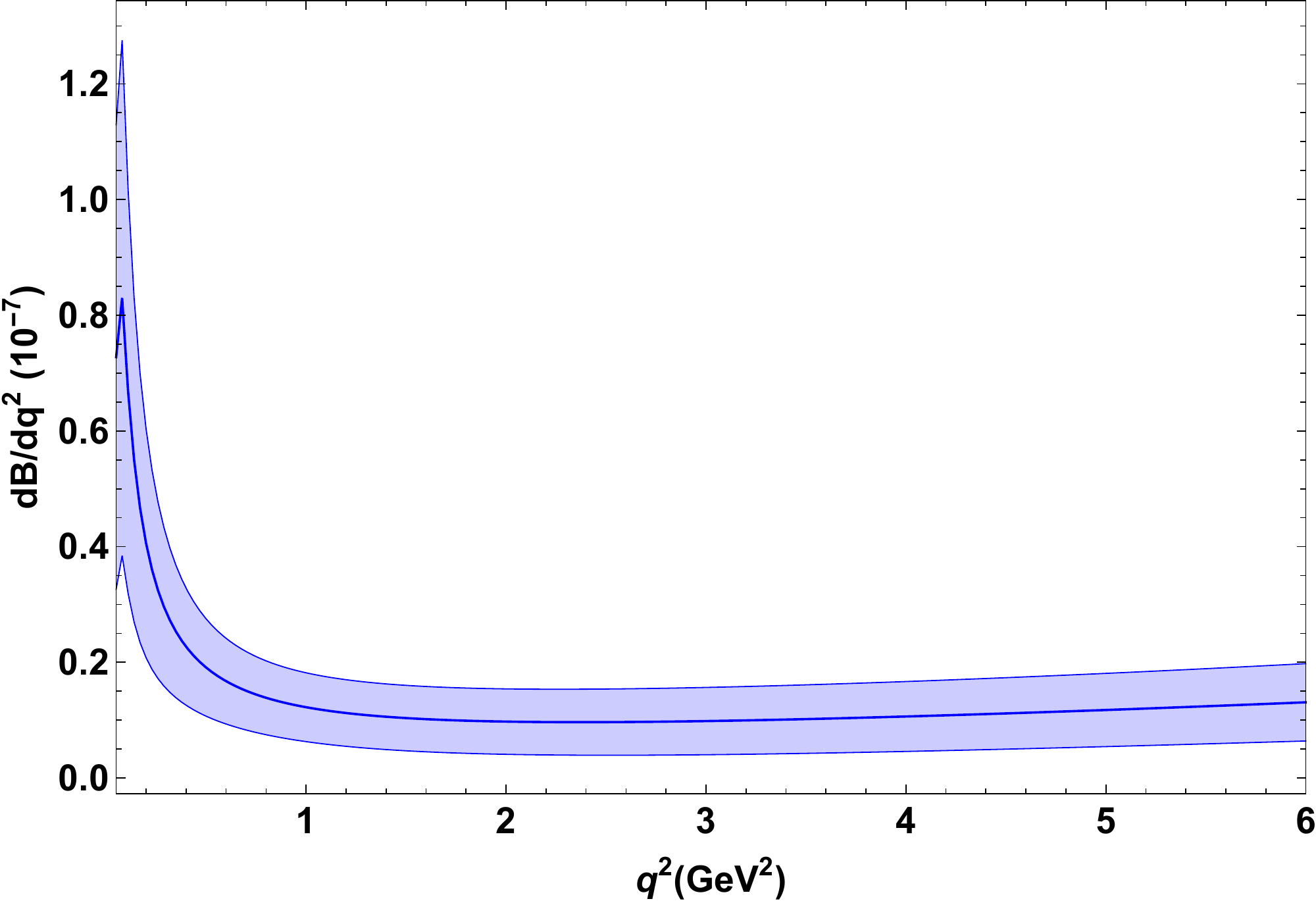}
		\caption{ Differential branching fraction $d{\cal B}/dq^2$,  forward-backward asymmetry of lepton pair, $A_{\rm FB}$, and the logitudinal polarization fraction, $F_L$,  as function of dimuon invariant mass $q^2$ for $B \to K_2^\ast \mu^+\mu^-$ in the SM.  The bands show estimates of uncertainties due to errors in form factors and various inputs (discussed in the  text)  used for the evaluation of observables.  }\label{SM_obs}
	\end{center}
\end{figure}

The study of the four-fold angular distribution gives access to numerous observables that can be measured by the LHCb. Due to factorization of long and short-distance physics at large recoil Eq.~(\ref{eq:low-recoil-Is}), one can construct observables in terms of ratios where the form factors cancel making them highly sensitive to NP. In the context of decay $B\to K^\ast(\to K \pi) \mu^+\mu^-$, such observables have been constructed (see, for example, \cite{Descotes-Genon:2013vna} and references therein). Taking cue from $B\to K^\ast\ell^+\ell^-$ \cite{Matias:2012xw,DescotesGenon:2012zf,Descotes-Genon:2013vna}, we consider following set of observables where the soft form factor cancel at leading order in $\alpha_s$ and $\Lambda_{\rm QCD}/m_b$ making them suitable probe for new physics 
\begin{eqnarray}
\langle P_1\rangle &=& \frac{1}{2} \frac{\displaystyle\int_{\rm bin} dq^2 I_3}{\displaystyle\int_{\rm bin}dq^2 I_2^s }\, ,\quad\quad  \langle P_4\rangle = \frac{\displaystyle\int_{\rm bin} dq^2 I_4}{\sqrt{-\displaystyle\int_{\rm bin}dq^2 I_2^c \displaystyle\int_{\rm bin}dq^2 I_2^s }}\, ,\\
\langle P_2\rangle &=& \frac{1}{8} \frac{\displaystyle\int_{\rm bin} dq^2 I_6 }{\displaystyle\int_{\rm bin} dq^2 I_2^s }\, ,\quad\quad  \langle P_5\rangle = \frac{\displaystyle\int_{\rm bin} dq^2 I_5}{2\sqrt{-\displaystyle\int_{\rm bin}dq^2 I_2^c \displaystyle\int_{\rm bin}dq^2 I_2^s }}\, ,\\
\langle P_3\rangle &=& -\frac{1}{4} \frac{\displaystyle\int_{\rm bin} dq^2 I_9}{\displaystyle\int_{\rm bin} dq^2 I_2^s }\, ,\quad\quad  \langle P_6\rangle = \frac{\displaystyle\int_{\rm bin} dq^2 I_7}{2\sqrt{-\displaystyle\int_{\rm bin}dq^2 I_2^c \displaystyle\int_{\rm bin}dq^2 I_2^s }}\, .
\end{eqnarray}

The subleading corrections to them will be estimated following the discussions in section~\ref{subsec:largeRec}. 
%In rest of the article, we will ignore superscript ``$K_2^\ast$" in $P_i^{(\prime) K_2^\ast}$, and will refer these as  $P_i^{(\prime)}$ for simplification of notations. 
%

As also discussed in the Sec.~\ref{sec:intro}, recent LHCb results \cite{Aaij:2014ora,Aaij:2017vbb} of measurements of the ratio of $B\to K(K^\ast)\ell^+\ell^-$ branching ratios of di-muon over di-electrons known as $R_{K,K^\ast}$ show significant deviation from their SM predictions $R_{K,K^\ast}\sim 1$ \cite{Bordone:2016gaq}, which hints to violation of lepton flavor universality. Observation of the same pattern of deviation in the $K$ and $K^\ast$ mode is quite intriguing and has attracted a lot of attention recently (see Ref.~\cite{Hiller:2017bzc} for a model-independent analysis) . If NP is to blame for these, such deviations should be seen in $B\to K^\ast_2\ell^+\ell^-$ as well, and need to be studied. We define similar ratio for $B\to K^\ast_2\ell^+\ell^-$
\begin{equation}
R_{K^\ast_2} = \frac{\mathcal{B}(B\to K^\ast_2\mu\mu)}{\mathcal{B}(B\to K^\ast_2ee)}\, .
\end{equation}

Having discussed all the observables, we are now ready to proceed with the numerical analysis of these observables in the SM and NP scenarios in the next section. 

%%%%%%%%%%%%%%%%%%%%%%%%%%%%%%%%%%%%%%%%%%%%%%%%%%%%%%%%%%%%%%%%%%%%%%%%%%%%%%%%%%%%%%%%%%%%
%%%%%%%%%%%%%%%%%%%%%%%%%%%%%%%%%%%%%%%%%%%%%%%%%%%%%%%%%%%%%%%%%%%%%%%%%%%%%%%%%%%%%%%%%%%%
\section{Numerical Analysis \label{sec:glfit}}
In the light of the recent flavor anomalies, several groups have performed global fits of Wilson coefficients to $b\to s\ell^+\ell^-$ data to decipher the pattern of NP \cite{Descotes-Genon:2013wba,Descotes-Genon:2015uva,Altmannshofer:2014rta,Hurth:2016fbr,Altmannshofer:2015sma,Altmannshofer:2013foa,Beaujean:2013soa,Hurth:2013ssa}. These fits indicate that a negative contribution to the Wilson coefficient $C_9^\mu$ can alleviate the tension between theory and experimental data. There are other scenarios as well which lead to similar fits. Following Ref.~\cite{Capdevila:2017bsm} we consider three of them (called S1, S2, S3) having largest pull\footnote{pull=$\sqrt{\Delta \chi^2}$} 
\begin{itemize}
\item S1: NP in $C_9$ only with $C_9^{\mu,\rm NP} = -1.1$, for which the pull is $5.8 \sigma$.

\item S2: In this scenario, NP is considered in both $C_9$ and $C_{10}$, but they are correlated, $C_9^{\mu,\rm NP}=-C_{10}^{\mu,\rm NP} = -0.6$ and the pull for this scenario is $5.3 \sigma$.

\item S3: In this scenario, NP is considered in $C_9$ and $C_9^{\prime}$ and again correlated with best fit given by $C_9^{\mu,\rm NP}= -C_9^{\prime~\mu,\rm NP}=-1.01$ for which the pull is $5.4\sigma$.
\end{itemize}

\begin{figure}[h!]
	\begin{center}\begin{tabular}{cc}
	\includegraphics[scale=0.4]{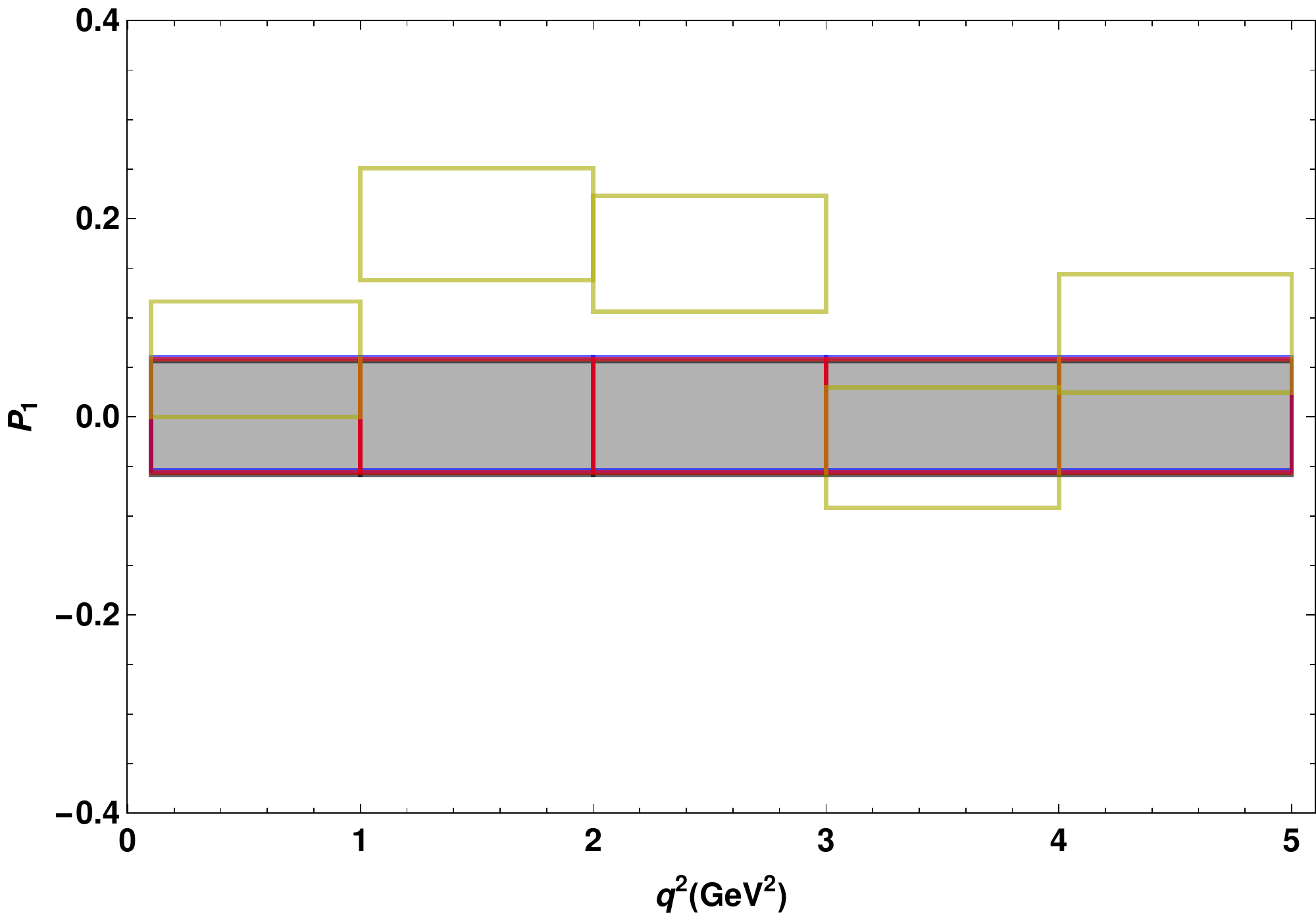}&
			\includegraphics[scale=0.4]{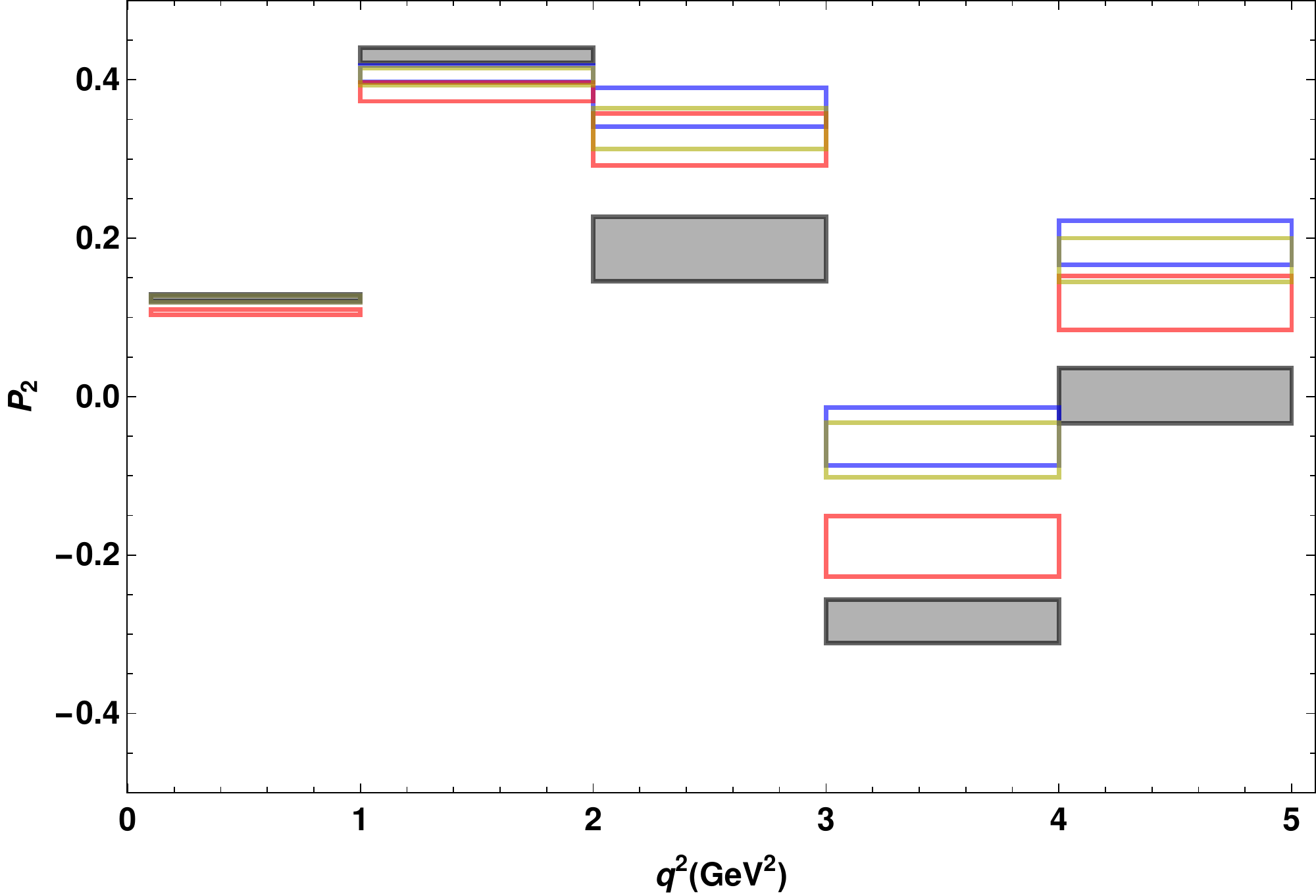}\\
			\includegraphics[scale=0.4]{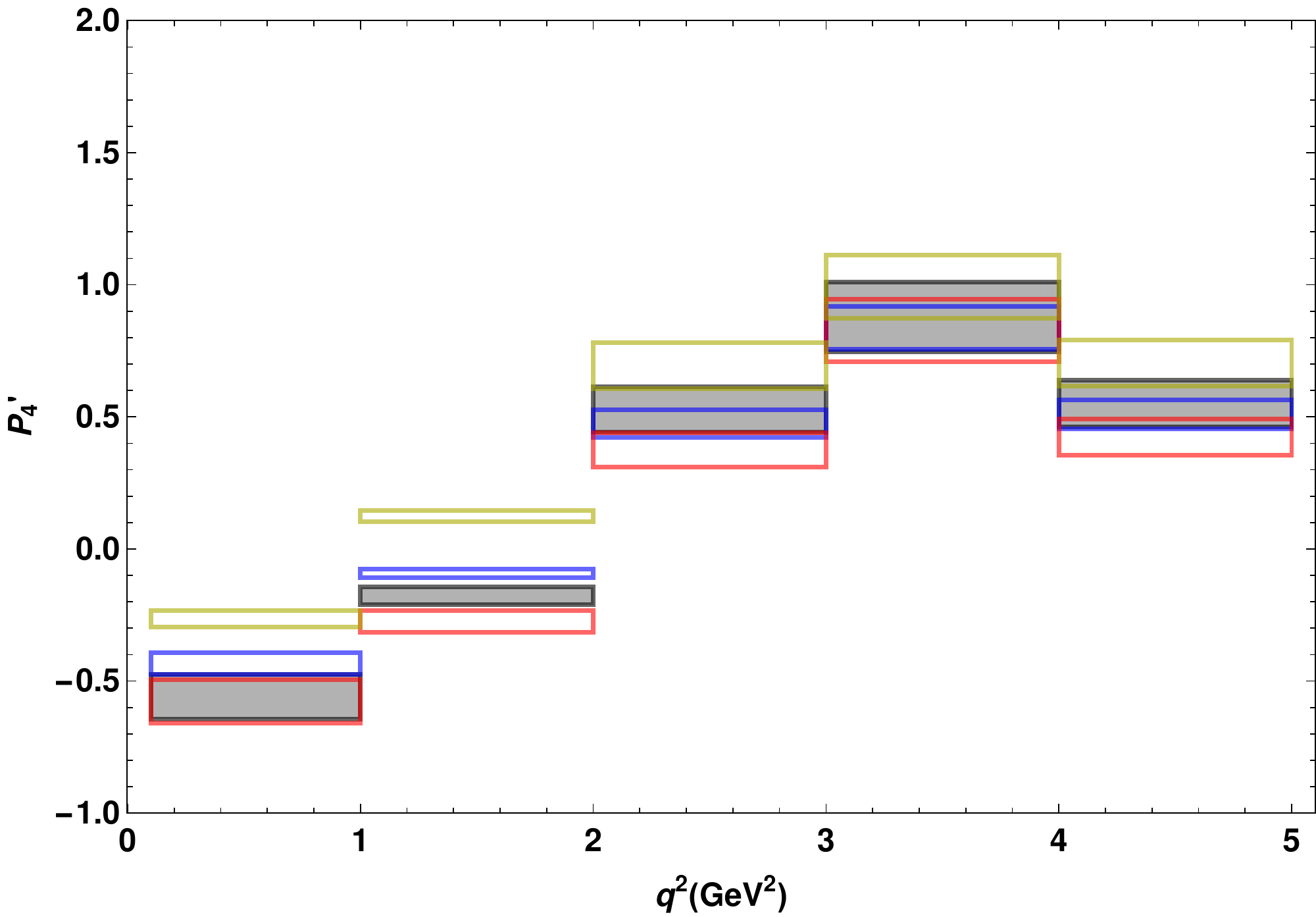}&
			\includegraphics[scale=0.4]{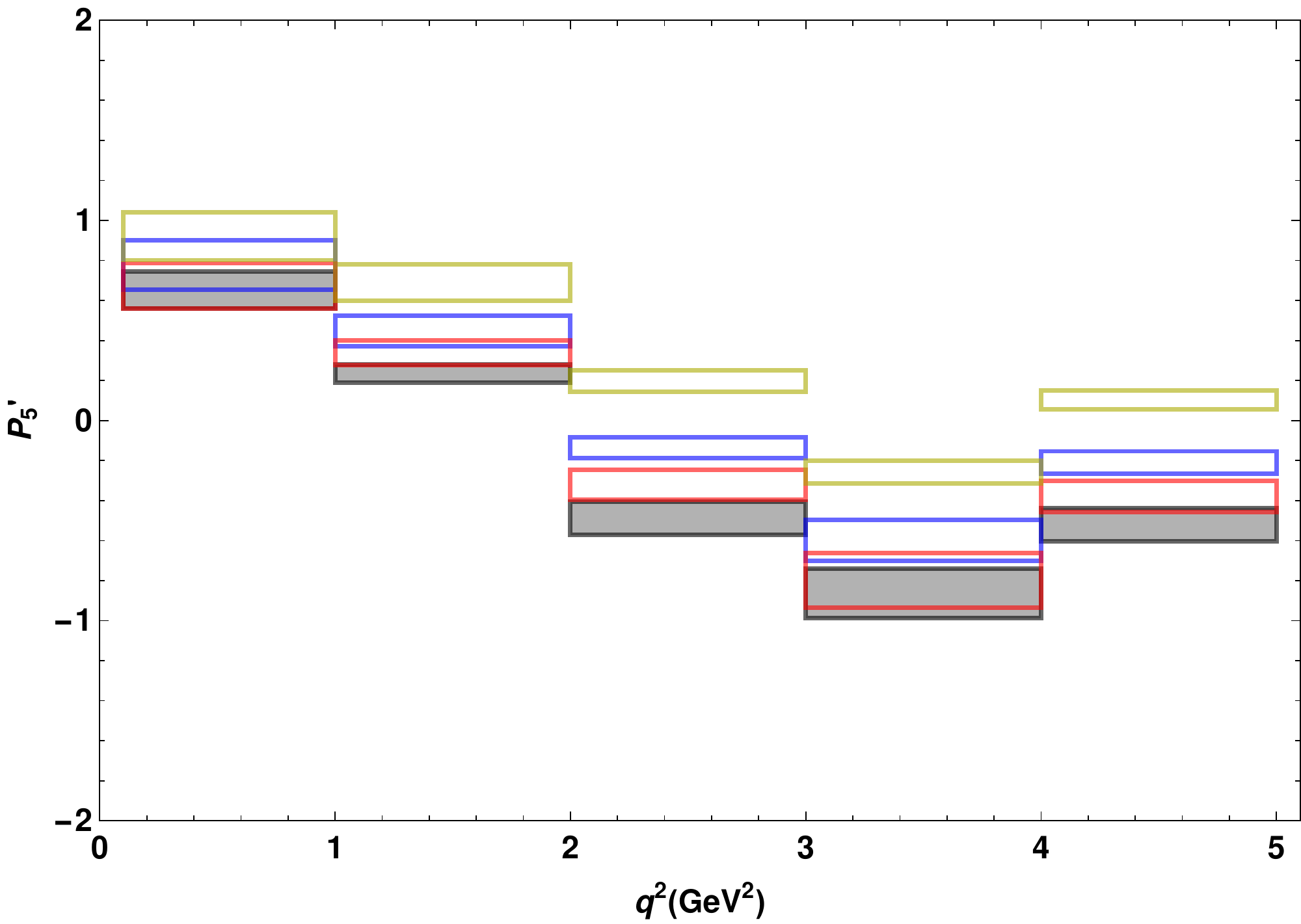}\\
			\includegraphics[scale=0.4]{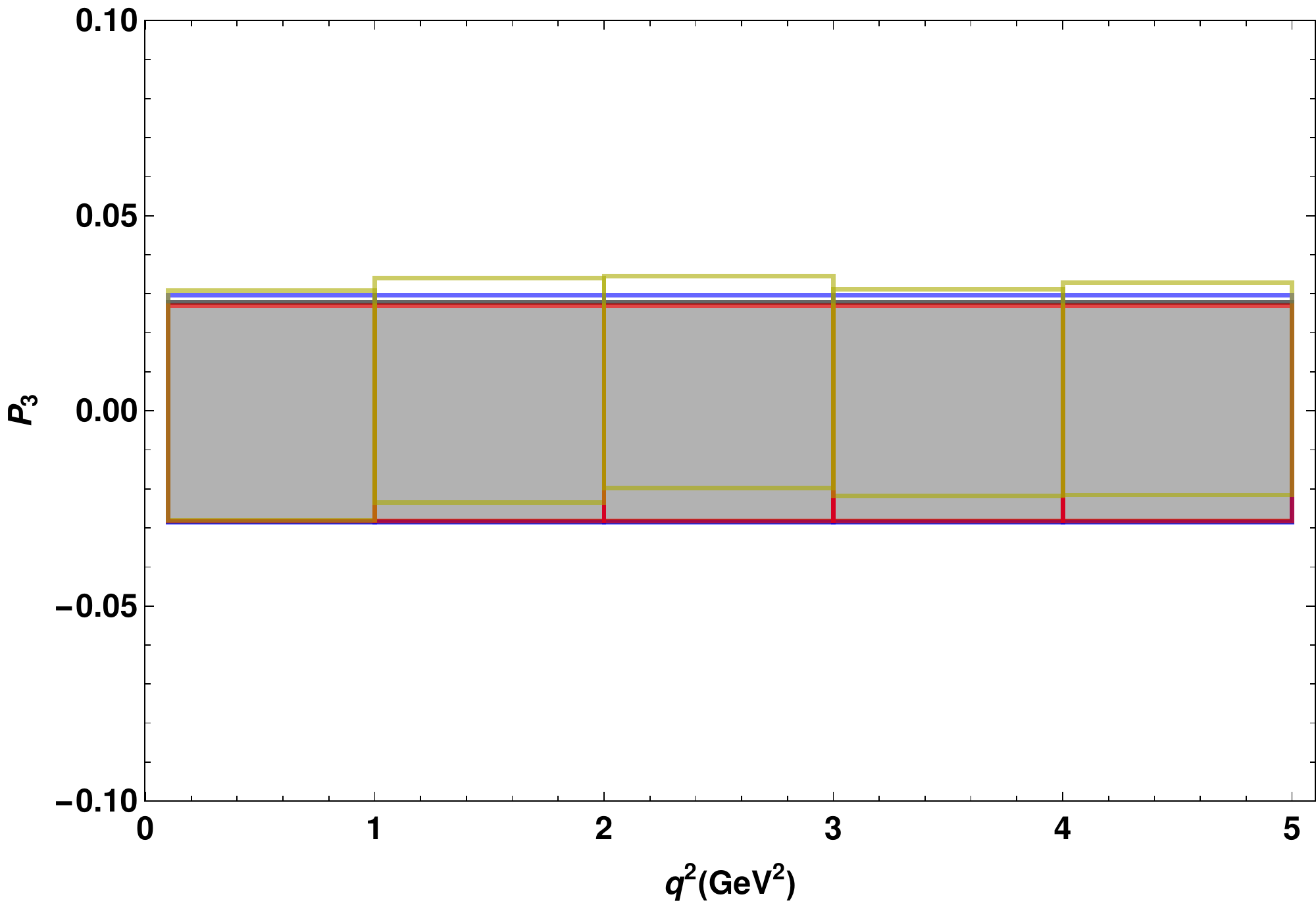}&
		\includegraphics[scale=0.4]{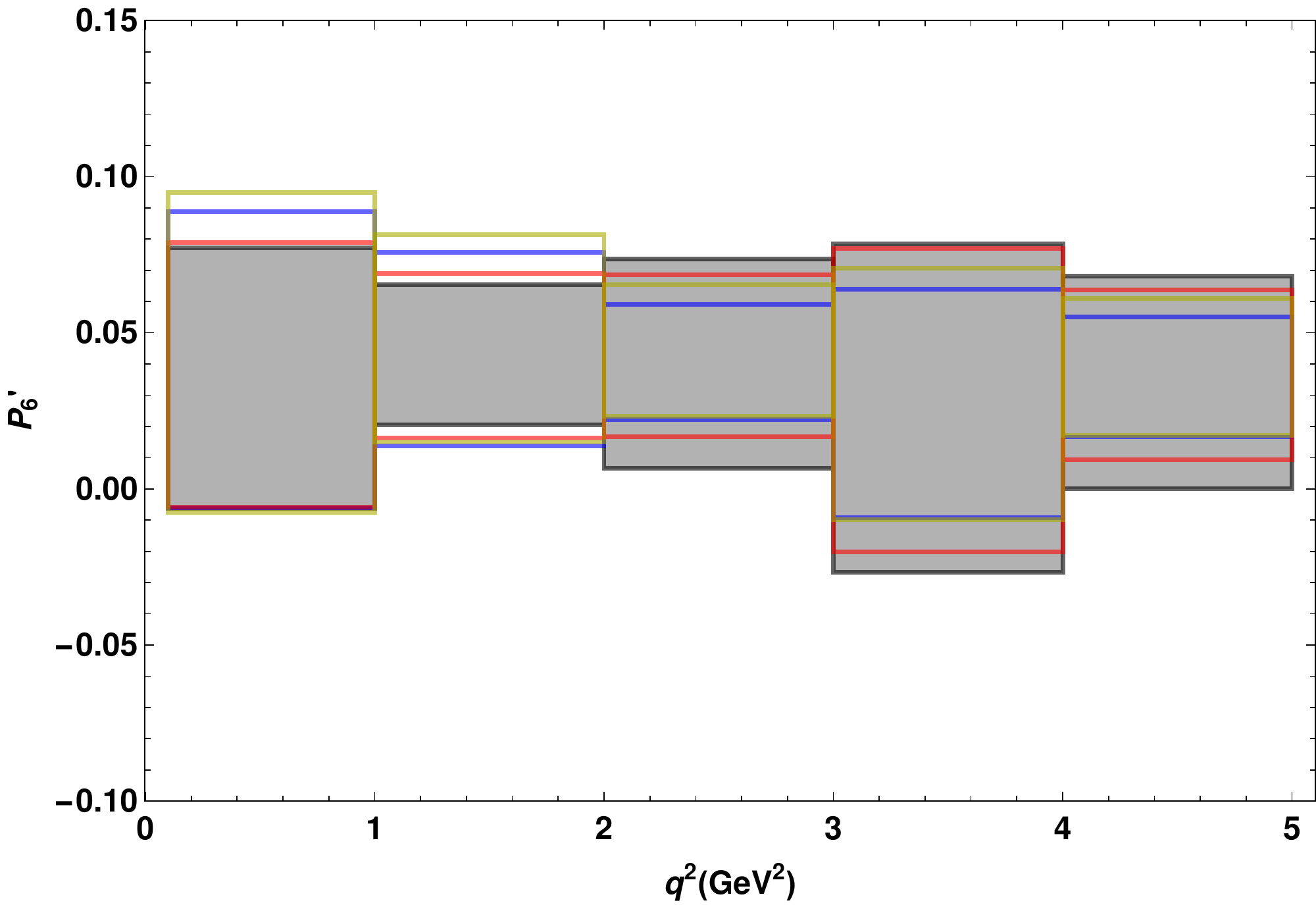}\\
		\end{tabular}
		\caption{The clean observables $P_i^{(\prime)}(q^2)~ (i=1,2,3,4,5,6)$ in different $q^2$ bins in the SM (grey) and three NP scenarios S1 (blue), S2 (red) and S3 (yellow).  The horizontal width of boxes corresponds to the $q^2$ bin size, and the vertical length gives estimate of uncertainties due to errors in form factors and other input for that particular $q^2$ bin.  }\label{binned_Ps}
	\end{center}
\end{figure}

Our main numerical results in the SM and the above three NP cases for all the angular observables considered in this work  are collected in Appendix~\ref{app:SMNPtables}.  The binned predictions for clean observables are displayed in Fig.~\ref{binned_Ps}. We  restrict our analysis to  low dilepton invariant mass region and  consider $q^2$ bins lying in range $0.1 - 6.0$ ${\rm GeV}^2$.

The branching fraction for $B\to K_2^\ast\mu^+\mu^-$ in the SM  is $\sim {\cal O}(10^{-7})$ (see Table~\ref{Table:normal_obs}). In all three NP scenarios, we find consistently smaller central values for branching fraction compared to the SM value. This is pertaining to the fact that the global analysis of $b\to s \ell^+\ell^-$ suggests destructive NP contribution to  $C_9^{\mu}$. For $A_{\rm FB}$ ($F_L$) we find slightly larger (smaller) central value in NP cases compared to the central SM value. However,  as these observables ($d\Gamma/dq^2$, $F_L$ and $A_{FB}$) are at present plagued by large theoretical errors,  no striking deviation from the SM value is found. On the other hand,  prospects for testing NP hypothesis in $b\to s \ell^+\ell^-$ in some clean observables $P_i(q^2)$ are promising.

The  clean observable $P_1$ depends on the angular coefficients $I_3$ and $I_2$, and is of special interest due to its remarkable sensitivity to right-handed currents. The $(V-A)$ structure of the SM renders the $H_\pm$ helicities of the $B\to K^\ast_2\ell^+\ell^-$  suppressed, implying $|A_\parallel| \simeq |A_\perp|$. Therefore, this observable is predicted to be zero in the SM. The similar charaterstic is  also shared by  its $B \to K^\ast$ counterpart as noted in \cite{Descotes-Genon:2015uva}. As shown in Fig.~\ref{binned_Ps}, $P_1$ is consistent with zero in the SM and in two scenarios S1 and S2 (which assume NP in the left-handed currents only), while a large  deviations from $P_1\simeq 0$ is found in scenario S3 (which has nonzero value of right-handed Wilson coefficient $C_9^\prime$). The observable $P_2$ is similar to forward-backward asymmetry $A_{\rm FB}$, but is theoretically much cleaner. Similar to $A_{\rm FB}$,  $P_2$ has larger values in all three NP scenarios.  The zero-crossing of $P_2$ (same as of $A_{FB}$\footnote{Note that since numerators of $P_2$ and $A_{FB}$ are same, the zeros of both observables are also same.}) lies in the [2,4] GeV$^2$ and at the leading order is given by
\bea
 q^2_0(P_2) \simeq -\frac{ 2\,C_7 }{C_9 - \left(C_{10}^\prime/C_{10}\right) C_9^\prime} m_b \,m_B.  
\eea 
In order to obtain the above relation, we have used transversity amplitudes given in Eqs.~\eqref{eq:TAsinSFF1}-\eqref{eq:TAsinSFF4} and assumed the Wilson coefficients to be real.    This expression is identical to the corresponding observable for $B\to K^\ast$ case.  Note that the zero crossing $q^2_0(P_2)$ depends on the short-distance Wilson coefficients $C_{9}^{\ell(\prime)}$ and $C_{10}^{\ell(\prime)}$, and has no dependence on the mass of lepton in final state. Consequently, in the SM it has  the same value for all three decay modes $B\to K_2^\ast \ell^+\ell^-$ ($\ell = e,\mu, \tau$).  Therefore  the zero crossing $q^2_0(P_2)$ turn out to be a good observable  to test the hypothesis of  lepton flavor universality violating (LFUV) NP.

For observables $P_4^\prime$ and $P_5^\prime$, the largest deviations from the SM value are observed in scenario S3, thereby showing good sensitivity to right-handed NP.  On the other hand, observables $P_3$ and $P_6^\prime$ depend on $I_9(q^2)$ and $I_7(q^2)$ respectively.   These two observables depend on imaginary part of  $\rho_{\mp}^{L,R}(q^2)$ and $\sigma_\mp(q^2)$.  The imaginary part of the SM Wilson coefficient $C_{9,7}^{\rm eff}$ is very tiny, and therefore the SM predictions for $P_3$ and $P_6^\prime$ are highly suppressed. Since, in our numerical analysis, we consider real NP Wilson coefficients, these observables remain suppressed in all three NP scenarios. Deviations in these observables, if seen in experiments, will be a sign of CP-violating NP.  

 \begin{figure}[h!]
	\begin{center}
		\includegraphics[scale=0.55]{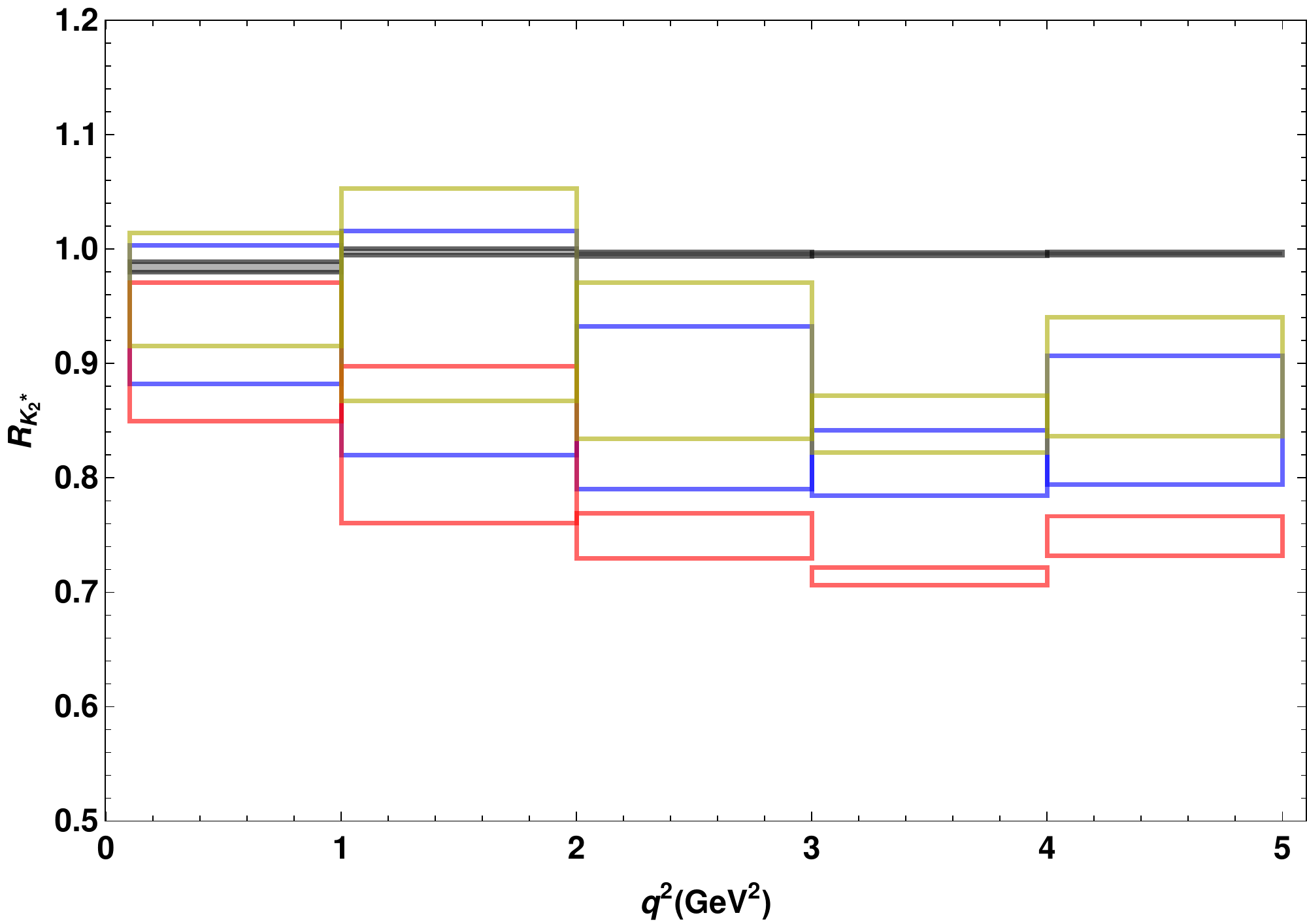}
				\caption{Binned predictions for $R_{K_2^\ast}$ in the SM (grey) and NP scenarios S1 (blue), S2 (red), and S3 (yellow).}\label{Fig:RK2}
	\end{center}
\end{figure}

Finally, in Fig.~\ref{Fig:RK2}} we present our determinations of the LFUV  ratio $R_{K_2^\ast}$.  Similar observables for $B\to K^{(\ast)}\ell^+\ell^-$ in the SM are predicted to be $\sim 1$ \cite{Bordone:2016gaq}. These LFUV ratios are exceptionally clean observables with theoretical errors being at the level of only $\sim 1 \%$, making them an ideal candidate to probe NP.  As mentioned earlier,  $R_K$ and $R_{K^\ast}$ have been measured experimentally and  both measurements are lower than  the SM value, which could be interpreted as sign of NP.  Therefore the measurement of $R_{K_2^\ast}$ can be important to corroborate the deviations seen in $R_K$ and $R_K^\ast$.  In all three NP scenarios, $R_{K_2^\ast}$ is suppressed compared to the SM value. For NP case S2, the deviations from unity are largest while for NP case S3 the suppression is relatively smaller as this solution contain a mixture of left-handed and  right-handed currents, and right-handed currents tend to increase the value of ratio.   The bin averaged predictions  for $R_{K_2}^\ast$ in the SM and NP cases are given  in Appendix~\ref{app:SMNPtables}.   %

 \begin{table}[H]
 	\begin{center}
 		\begin{tabular}{c|c|c}
 			\hline\hline
 			Parameter           &     Value & Source \\\hline
 			$m_{B}$ & $ 5.279$ GeV & \cite{Tanabashi:2018oca}\\
 			$m_{K_2^\ast}$ & $1432.4\pm 1.3\,\text{MeV}$  & \cite{Tanabashi:2018oca}\\
 			$m_b^{\overline{\rm MS}}$ & 4.20~{\rm GeV}& \cite{Detmold:2016pkz}\\
 			$m_b^{\text{pole}}$ & $4.7417$ GeV & \cite{Detmold:2016pkz}\\
 			$m_c^{\text{pole}}$ & $1.5953$ GeV & \cite{Detmold:2016pkz}\\
 			$|V_{ts}^* V_{tb}|$ & $0.04088 \pm 0.00055$ &  \cite{utfit}\\
 			$\alpha_s(\mu = 4.2~{\rm GeV})$ &0.2233 & \cite{Detmold:2016pkz} \\
 			$\alpha_e(\mu = 4.2~{\rm GeV})$ & 1/133.28 & \cite{Detmold:2016pkz} \\
			$Br(K_2^\ast \to K \pi)$ & $(49.9 \pm 1.2)\%$ &\cite{Tanabashi:2018oca}\\\hline\hline
 		\end{tabular}
 	\end{center}
 	\caption{ The numerical inputs used in our analysis. The values of  $\alpha_s$, $\alpha_e$, and $m_b^{\overline{\rm MS}}$ at low scale $\mu=2.1~{\rm GeV}$ and high scale $\mu=8.4~{\rm GeV}$ are also used from Ref.~\cite{Detmold:2016pkz}.}
 	\label{tab:input}
 \end{table}

 \begin{table}[H]\begin{center}
	\setlength{\tabcolsep}{5pt}
	\begin{tabular}{ c| c | c | c | c | c | c | c | c | c | c  }
		\hline\hline
		&$C_1$  & $C_2$ & $C_3$ & $C_4$ & $C_5$ &$C_6$ & $C_7$ & $C_8$ & $C_9$ & $C_{10}$  \\
		\hline
		$\mu=2.1$ GeV&$-0.4965$ & $1.0246$ & $-0.0143$ & $-0.1500$ & $0.0010$ & $0.0032$ & $-0.3782$ & $-0.2133$ & $4.5692$ & $-4.1602$\\
		\hline
		$\mu=4.2$ GeV&$-0.2877$ & $1.0101$ & $-0.0060$ & $-0.0860$ & $0.0004$ & $0.0011$ & $-0.3361$ & $-0.1821$ & $4.2745$ & $-4.1602$\\
		\hline
		$\mu=8.4$ GeV&$-0.1488$ & $1.0036$ & $-0.0027$ & $-0.0543$ & $0.0002$ & $0.0004$ & $-0.3036$ & $-0.1629$ & $3.8698$ & $-4.1602$\\
		\hline\hline
	\end{tabular}
	 \end{center}
	\caption{Values of SM Wilson coefficients taken from Ref.~\cite{Detmold:2016pkz}}  
	\label{Table:WC}
\end{table}
%%%%%%%
 %%%%%
%\section{Numerical Analysis}
%%%%%%%%%%%%%%%%%%%%%%%%%%%%%%%%%%%%%%%%%%%%%%%%%%%%%%%%%%%%%%%%%%

\section{Summary and Discussion \label{Sec:summary}}
In this paper, we have performed an angular analysis of exclusive semileptonic decay $B\to K_2^\ast(\to K \pi)\mu^+\mu^-$. The decay, at the quark level, is governed by the $b\to s \ell^+\ell^-$ FCNC transition. About $2-3~\sigma$ discrepancies in $b\to s\ell^+\ell^-$ transitions have recently been observed in $B \to K(K^\ast)\ell^+\ell^-$ decays. If these discrepancies are due to NP then similar anomalies are also expected in  $B\to K_2^\ast(\to K \pi)\mu^+\mu^-$ transitions as well which make this decay worth studying.

A full angular distribution of $B\to K_2^\ast(\to K \pi)\mu^+\mu^-$ in the transversity basis, similar to $B\to K^\ast(\to K \pi)\mu^+\mu^-$ offers a large number of observables. We have worked in the limit of heavy quark $m_b\to \infty$ and large energy $E_{K_2^\ast}\to \infty$ where symmetry relations reduce the number of independent form factors from seven to two: $\xi_\perp (q^2)$ and $\xi_\parallel (q^2)$. Utilising these symmetry relations we have provided expressions for transversity amplitudes, and have constructed new clean angular observables. The form factor dependence for these clean observables cancel at leading order in $\alpha_s$ and $\Lambda_{\rm QCD}/m_b$. The uncertainties due to the sub-leading corrections have been included in our numerical analysis.

We have presented determinations of $B\to K_2^\ast(\to K \pi)\mu^+\mu^-$ decay rate, forward-backward asymmetry, longitudinal polarization fractions and clean observables in the SM and several NP cases . The NP scenarios are motivated by the recent global fits to the $b\to s\ell^+\ell^-$ data. We have also considered the LFU violation sensitive observable $R_{K_2^\ast}$. The $B\to K_2^\ast(\to K \pi)\mu^+\mu^-$ decay may provide new and complementary information to $B\to K^\ast (K)\mu^+\mu^-$ in searches of NP

%%%%%%%%%%%%%%%%%%%%%%%%%%%%%%%%%%%%%%%%%%%%%%%%%%%%%%%%%%%%%%%%%%%%%%
  \begin{acknowledgements}
  	DD would like to thank DST, Govt. of India for the financial support under INSPIRE Faculty Fellowship. Authors would like to thank Gagan Mohanty and Saurabh Sandilya for useful discussions. 
  	  \end{acknowledgements}

\appendix
%%%%%%%%%%%%%%%%%%%%%%%%%%%%%%%%%%%%%%%%%%%%%%%%%%%%%%%%%%%%%%%%%%%
%%%%%%%%%%%%%%%%%%%%%%%%%%%%%%%%%%%%%%%%%%%%%%%%%%%%%%%%%%%%%%%%%%%
%%%%%%%%%%%%%%%%%%%%%%%%%%%%%%%%%%%%%%%%%%%%%%%%%%%%%%%%%%%%%%%%%%%%%%%%%%%%%%%%%%%%%%%%%%%%%%%%%%%%%%%%%%%%%%%%%%%%%%%%%%%%%%%%%%%%%%%%%%%%%%%%%%%%%%%%%%%%%%%%%%%%%%%%%%%
\section{Effective Wilson coefficients for $b\to s\ell^+\ell^-$ transition \label{app:C79eff}}
Corresponding to the $b\to s\ell^+\ell^-$ effective Hamiltonian equation~(\ref{eq:Heff}), the one loop contributions from $\mathcal{O}_1-\mathcal{O}_6$ to the $\mathcal{O}_7$ and $\mathcal{O}_9$ are absorbed by defining the effective Wilson coefficients $C_7^{\rm eff}$ and $C_9^{\rm eff}$ \cite{Detmold:2016pkz}
\begin{equation}
\begin{split}
C_7^{\rm eff}(\mu) &= C_7 - \frac{1}{3}C_3 + \frac{4}{3}C_4 + 20C_5 + \frac{80}{3}C_6 -\frac{\alpha_S}{4\pi}\left[(C_1-6C_2)F_{1,c}^{(7)}(q^2)+C_8F_8^{(7)}(q^2)\right] ,\\
C_9^{\rm eff}(\mu) &= C_9 + h(q^2, m_c) \Big(\frac{4}{3} C_1 + C_2 + 6 C_3 + 60 C_5 \Big)\, \\
&-\frac{1}{2} h(q^2, m_b) \Big( 7 C_3 + \frac{4}{3} C_4 + 76 C_5 + \frac{64}{3} C_6 \Big)  - \frac{1}{2} h(q^2,0) \Big( C_3 + \frac{4}{3} C_4 + 16 C_5 + \frac{64}{3} C_6 \Big)\, \\
& + \frac{4}{3} C_3 + \frac{64}{9} C_5  + \frac{64}{27} C_6 -\frac{\alpha_S}{4\pi}\left[C_1F_{1,c}^{(9)}(q^2)+C_2F_{2,c}^{(9)} (q^2)+C_8F_8^{(9)}(q^2)\right].
\end{split}
\end{equation}
The value of the SM Wilson coefficients $C_i$ $(i=1, 2,..10)$ are given in Table~\ref{Table:WC}.  The functions $h(q^2, m_q)$ and $F_8^{(7,9)}(q^2)$ can be found in Ref.~\cite{Beneke:2001at}, and the functions $F_{1,2}^{(7,9)}(q^2)$ are taken from Ref.~\cite{Asatryan:2001zw}. The values of masses of charm and bottom quark in these expressions are defined in pole mass scheme and are given in Table \ref{tab:input}.

\section{$K_2^\ast$ polarization tensors \label{app:polz}}
The tensor meson $K_2^\ast$ is described in terms of spin-2 polarization tensor $\epsilon^{\mu\nu}(n)$, where the
helicity $n$ can be $\pm 2, \pm 1$ and 0. The polarization tensor satisfy $\epsilon^{\mu\nu} k^\nu = 0$.
For the $K_2^\ast$ which has four momentum $(k_0, 0, 0, \vec{k})$, the polarization tensor $\epsilon^{\mu\nu}(h)$ can be constructed in terms of following polarization tensors \cite{Berger:2000wt}
\begin{equation}
\epsilon_\mu(0) = \frac{1}{m_{K_2^\ast}}(|\vec{k}|,0,0,k_0)\, ,\quad \epsilon_\mu(\pm) = \frac{1}{\sqrt{2}}(0,\mp 1, -i, 0)\, ,\nn
\end{equation}
in the following way
\begin{align}
& \epsilon_{\mu\nu}(\pm 2) = \epsilon_\mu(\pm 1)\epsilon_\nu(\pm 1)\, ,\quad
\epsilon_{\mu\nu}(\pm 1) = \frac{1}{\sqrt{2}}\Big[\epsilon_\nu(\pm)\epsilon_\nu(0) + \epsilon_\nu(\pm)\epsilon_\mu(0) \Big]\, ,\nn\\
& \epsilon_{\mu\nu}(0) = \frac{1}{\sqrt{6}}\Big[\epsilon_\mu(+) \epsilon_\nu(-) + \epsilon_\nu(+) \epsilon_\mu(-) \Big]
+ \sqrt{\frac{2}{3}}\epsilon_\mu(0)\epsilon_\nu(0)\, .\nn
\end{align}
In the decay under consideration, since there are two leptons in the final state the $n=\pm 2$ helicity states of the $K_2^\ast$
is not realized. It is therefore convenient to introduce a new polarization vector \cite{Wang:2010ni}
\begin{equation}
\epsilon_{T\mu}(h) = \frac{\epsilon_{\mu\nu}p^\nu}{m_B}\, ,\nn
\end{equation}
where $p$ is the four momentum of $B$ meson. The explicit expressions of polarization vectors are
\begin{eqnarray}
\epsilon_{T\mu}(\pm 1) &=& \frac{1}{m_B}\frac{1}{\sqrt{2}}\epsilon(0).p\, \epsilon_\mu(\pm) = \frac{\sqrt{\lambda}}{\sqrt{8}m_B m_{K_2}^\ast} \epsilon_\mu(\pm)\, ,\\
\epsilon_{T\mu}(0) &=& \frac{1}{m_B}\sqrt{\frac{2}{3}}\epsilon(0).p \,\epsilon_\mu(0) = \frac{\sqrt{\lambda}}{\sqrt{6}m_B m_{K_2}^\ast} \epsilon_\mu(0)\, ,
\end{eqnarray}
where $\lambda = m_B^4+m_{K_2^\ast}^4+q^4-2(m_B^2 m_{K_2^\ast}^2 +m_B^2 q^2 + m_{K_2^\ast}^2q^2)$.
\section{Angular Coefficients $I_i(q^2)$ \label{app:Is}}
Here we summarize the expressions of the angular coefficients appearing in the differential decay rate eq.~(\ref{eq:diff4}) in terms of transversity amplitudes \cite{Li:2010ra}
\begin{eqnarray}
I_1^c&=&  (|A_{0L}|^2+|A_{0R}|^2)
+8 \frac{m_\ell^2}{q^2}{\rm Re}[A_{0L}A^*_{0R} ]+4\frac{m_\ell^2}{q^2} |A_t|^2, \nonumber\\
I_1^s&=&\frac{3}{4} \left(1-\frac{4m_\ell^2}{3q^2}\right)[|A_{\perp L}|^2+|A_{||L}|^2+|A_{\perp R}|^2+|A_{||R}|^2  ]
+\frac{4m_\ell^2}{q^2} {\rm Re}[A_{\perp L}A_{\perp R}^*
+ A_{||L}A_{||R}^*],\nonumber\\
I_2^c  &=& -\beta_\ell^2(  |A_{0L}|^2+ |A_{0R}|^2)\,,\quad
I_2^s  =
\frac{1}{4}\beta_l^2(|A_{\perp L}|^2+|A_{||L}|^2+|A_{\perp R}|^2+|A_{||R}|^2),
\nonumber\\
I_3  &=&\frac{1}{2}\beta_l^2(|A_{\perp L}|^2-|A_{||L}|^2+|A_{\perp R}|^2-|A_{||R}|^2)\,\\
I_4
&=& \frac{1}{\sqrt2}\beta_\ell^2
[{\rm Re}(A_{0L}A_{||L}^*)+{\rm
	Re}(A_{0R}A_{||R}^*]\, ,\quad
I_5
= \sqrt 2\beta_l
[{\rm Re}(A_{0L}A_{\perp L}^*)-{\rm Re}(A_{0R}A_{\perp R}^*)],\nonumber\\
I_6  &=& 2\beta_\ell 
[{\rm Re}(A_{||L}A^*_{\perp L})-{\rm
	Re}(A_{||R}A^*_{\perp R})]\, ,\quad
I_7
= \sqrt2\beta_\ell
[{\rm Im}(A_{0L}A^*_{||L})-{\rm Im}(A_{0R}A^*_{||R})],\nonumber\\
I_8 &=& \frac{1}{\sqrt2}\beta_\ell^2
[{\rm Im}(A_{0L}A^*_{\perp L})+{\rm
	Im}(A_{0R}A^*_{\perp R})]\, ,\quad 
I_9
=\beta_\ell^2 
[{\rm Im}(A_{||L}A^*_{\perp L})+{\rm
	Im}(A_{||R}A^*_{\perp R})],\nn
\end{eqnarray}
where $\beta_\ell = \sqrt{1-\frac{4 m^2_\ell}{q^2}}$.

%%%%%%%%%%%%%%%%%%%%%%%%%%%%%%%%%%%%%%%%%%%%%%%%%%%%%%%%%%%%%%%%%%
\section{ \label{app:SMNPtables}}
\subsection{Prediction of observables in the SM}
\begin{longtable}{@{}lcrr@{}}
\toprule[1.6pt] 
 \hspace{0mm} Bin \hspace{10mm} & $P_1$\hspace{22mm} & $P_2$\hspace{24mm} & $P_3$\hspace{10mm}  \\ 
 \midrule 
 $ [0.1,1]$ \hspace{10mm} & $0.001\pm 0.059$\hspace{14mm} & $0.125\pm 0.004$\hspace{14mm} & $0.0\pm 0.028$ \\ 
 $ [1,2] $ & $0.001\pm 0.059$\hspace{14mm} & $0.431\pm 0.009$\hspace{14mm} & $0.0\pm 0.028$ \\ 
 $ [2,4] $ & $0.001\pm 0.059$\hspace{14mm} & $0.186\pm 0.041$ \hspace{14mm} & $0.0\pm 0.028$ \\ 
 $ [4,6] $ & $0.001\pm 0.059$\hspace{14mm} & $-0.284\pm 0.028$ \hspace{14mm} & $0.0\pm 0.028$ \\ 
 $ [1,6] $ & $0.001\pm 0.059$\hspace{14mm} & $0.001\pm 0.035$ \hspace{14mm} & $0.0\pm 0.028$ \\ 
\midrule[1.6pt] 
 \hspace{0mm} Bin & $P_4^\prime$\hspace{22mm} & $P_5^\prime$\hspace{24mm} & $P_6^\prime$\hspace{10mm}  \\ 
 \midrule 
$ [0.1,1] $ & $-0.560\pm 0.084$\hspace{14mm} & $0.652\pm 0.093$\hspace{14mm} & $0.035\pm0.042$ \\ 
 $ [1,2] $ & $-0.178\pm 0.034$\hspace{14mm} & $0.235\pm 0.0.045$\hspace{14mm} & $0.043\pm 0.023$ \\ 
 $ [2,4] $ & $0.527\pm 0.086$\hspace{14mm} & $-0.489\pm 0.083$\hspace{14mm} & $0.040\pm 0.034$ \\ 
 $ [4,6] $ & $0.878\pm 0.132$\hspace{14mm} & $-0.864\pm 0.124$\hspace{14mm} & $0.026\pm 0.053$ \\ 
 $ [1,6] $ & $0.550\pm 0.087$\hspace{14mm} & $-0.520\pm 0.083$\hspace{14mm} & $0.034\pm 0.034$ \\  
\midrule[1.6pt] 
\hspace{0mm} Bin  & $BR~(10^{-7})$\hspace{22mm} & $A_{FB}$\hspace{24mm} & $F_L $\hspace{10mm}  \\ 
\midrule 
$ [0.1,1] $ & $0.205\pm0.093$\hspace{14mm} & $0.094\pm0.029$\hspace{14mm} & $0.346\pm0.198$ \\ 
$ [1,2] $ & $0.104\pm0.055$\hspace{14mm} & $0.202\pm0.133$\hspace{14mm} & $0.686\pm0.208$ \\ 
$ [2,4] $ & $0.196\pm0.119$\hspace{14mm} & $0.070\pm0.059$\hspace{14mm} & $0.760\pm0.191$ \\ 
$ [4,6] $ & $0.232\pm0.122$\hspace{14mm} & $-0.141\pm0.094$\hspace{14mm} & $0.679\pm0.210$ \\ 
$ [1,6] $ & $0.532\pm0.289$\hspace{14mm} & $0.0\pm0.019$\hspace{14mm} & $0.709\pm0.204$ \\  
\bottomrule[1.6pt] 
\end{longtable}
%%%%
\subsection{Prediction of observables in the NP scenario S1 ($C_9^{\mu, {\rm NP}} = -1.1$) }
\begin{longtable}{@{}lcrr@{}}
	\toprule[1.6pt] 
	\hspace{0mm} Bin \hspace{10mm} & $P_1$\hspace{22mm} & $P_2$\hspace{24mm} & $P_3$\hspace{10mm}  \\ 
	\midrule 
	$ [0.1,1] $ \hspace{10mm} & $0.003\pm 0.057$\hspace{14mm} & $0.123\pm 0.004$\hspace{14mm}  & $ 0.0\pm 0.029 $ \\ 
	$ [1,2] $ & $0.003\pm 0.057$\hspace{14mm} & $0.408\pm 0.011$ \hspace{14mm} & $0.0\pm 0.029$ \\ 
	$ [2,4] $ & $0.003\pm 0.057$\hspace{14mm} &  $0.365\pm 0.024$ \hspace{14mm} & $ 0.0\pm 0.029 $ \\ 
	$ [4,6] $ & $0.003\pm 0.057$\hspace{14mm} &  $-0.050\pm 0.037$ \hspace{14mm}  & $ 0.0\pm 0.029$ \\ 
	$ [1,6] $ & $0.003\pm 0.057$\hspace{14mm} & $0.194\pm 0.028$ \hspace{14mm} & $ 0.0\pm 0.029 $ \\ 
	\midrule[1.6pt] 
	\hspace{0mm} Bin \hspace{10mm} & $P_4^\prime$\hspace{22mm} & $P_5^\prime$\hspace{24mm} & $P_6^\prime$\hspace{10mm}  \\ 
	\midrule 
	$ [0.1,1] $\hspace{10mm} & $-0.434\pm 0.041$\hspace{14mm} & $0.777\pm 0.124$\hspace{14mm} & $0.041\pm 0.047$ \\ 
	$ [1,2] $ & $-0.093\pm 0.016$\hspace{14mm} & $0.448\pm 0.076$\hspace{14mm} & $0.048\pm 0.031$ \\ 
	$ [2,4] $ & $0.474\pm 0.053$\hspace{14mm} & $-0.136\pm 0.052$\hspace{14mm} & $0.041\pm 0.018$ \\ 
	$ [4,6] $ & $0.837\pm 0.081$\hspace{14mm} & $-0.597\pm 0.102$\hspace{14mm} & $0.027\pm 0.037$ \\ 
	$ [1,6] $ & $0.509\pm 0.054$\hspace{14mm} & $-0.210\pm 0.056$\hspace{14mm} & $0.036\pm 0.019$ \\  
	\midrule[1.6pt] 
	\hspace{0mm} Bin & $BR~(10^{-7})$\hspace{22mm} & $A_{FB}$\hspace{24mm} & $F_L $\hspace{10mm}  \\ 
	\midrule 
	$ [0.1,1]$& $0.201\pm0.094$\hspace{14mm} & $0.100\pm0.026$\hspace{14mm} & $0.288\pm0.188$ \\ 
	$ [1,2] $ & $0.093\pm0.046$\hspace{14mm} & $0.240\pm0.130$\hspace{14mm} & $0.594\pm0.222$ \\ 
	$ [2,4] $ & $0.165\pm0.090$\hspace{14mm} & $0.170\pm0.115$\hspace{14mm} & $0.691\pm0.210$ \\ 
	$ [4,6] $ & $0.188\pm0.097$\hspace{14mm} & $-0.027\pm0.028$\hspace{14mm} & $0.644\pm0.218$ \\ 
	$ [1,6] $ & $0.446\pm0.232$\hspace{14mm} & $0.102\pm0.065$\hspace{14mm} & $0.650\pm0.217$ \\  
	\bottomrule[1.6pt] 
\end{longtable}
%%%%
\subsection{Prediction of observables in the NP scenario S2 ($C_9^{\mu, {\rm NP}} = - C_{10}^{\mu,{\rm NP}} =-0.6$) }
\begin{longtable}{@{}lrrr@{}}
	\toprule[1.6pt] 
	\hspace{0mm} Bin \hspace{10mm} & $P_1$\hspace{22mm} & $P_2$ \hspace{24mm} & $P_3$\hspace{10mm}  \\ 
	\midrule 
	$ [0.1,1] $ \hspace{10mm}  & $0.0\pm 0.057$\hspace{14mm} & $0.107\pm 0.004$ \hspace{14mm}  & $ 0.0\pm 0.028 $\hspace{1mm} \\ 
	$ [1,2] $ & $0.0\pm 0.057$\hspace{14mm} &  $0.384 \pm 0.012$ \hspace{14mm} & $ 0.0\pm 0.028$ \hspace{6mm}\\ 
	$ [2,4] $ & $0.0\pm 0.057$\hspace{14mm} &  $0.325 \pm 0.033$ \hspace{14mm} & $0.0\pm 0.028 $ \hspace{6mm}\\ 
	$ [4,6] $ & $0.0\pm 0.057$\hspace{14mm} &  $-0.189\pm 0.038$ \hspace{14mm}  & $ 0.0\pm 0.028 $ \hspace{6mm}\\ 
	$ [1,6] $ & $0.0\pm 0.057$\hspace{14mm} &  $0.118 \pm 0.034$ \hspace{14mm} & $ 0.0\pm 0.028 $ \hspace{6mm}\\ 
	\midrule[1.6pt] 
	\hspace{0mm} Bin \hspace{10mm} & $P_4^\prime$\hspace{22mm} & $P_5^\prime$\hspace{24mm} & $P_6^\prime$\hspace{10mm}  \\ 
	\midrule 
	$ [0.1,1] $ & $-0.577\pm 0.082$\hspace{14mm} & $0.673\pm 0.118$\hspace{14mm} & $0.037\pm 0.042$ \\ 
	$ [1,2] $ & $-0.275\pm 0.042$\hspace{14mm} & $0.339\pm 0.064$\hspace{14mm} & $0.043\pm 0.027$ \\ 
	$ [2,4] $ & $0.374\pm 0.065$\hspace{14mm} & $-0.321\pm 0.139$\hspace{14mm} & $0.043\pm 0.026$ \\ 
	$ [4,6] $ & $0.826\pm 0.118$\hspace{14mm} & $-0.899\pm 0.183$\hspace{14mm} & $0.028\pm 0.049$ \\ 
	$ [1,6] $ & $0.423\pm 0.068$\hspace{14mm} & $-0.582\pm 0.131$\hspace{14mm} & $0.037\pm 0.027$ \\  
	\midrule[1.6pt] 
	\hspace{0mm} Bin & $BR~(10^{-7})$\hspace{22mm} & $A_{FB}$\hspace{24mm} & $F_L $\hspace{10mm}  \\ 
	\midrule 
	$ [0.1,1]$& $0.189\pm0.091$\hspace{14mm} & $0.087\pm0.023$\hspace{14mm} & $0.284\pm0.186$ \\ 
	$ [1,2] $ & $0.083\pm0.042$\hspace{14mm} & $0.213\pm0.119$\hspace{14mm} & $0.616\pm0.219$ \\ 
	$ [2,4] $ & $0.144\pm0.084$\hspace{14mm} & $0.128\pm0.097$\hspace{14mm} & $0.738\pm0.198$ \\ 
	$ [4,6] $ & $0.165\pm0.091$\hspace{14mm} & $-0.090\pm0.063$\hspace{14mm} & $0.685\pm0.209$ \\ 
	$ [1,6] $ & $0.393\pm0.217$\hspace{14mm} & $0.055\pm0.041$\hspace{14mm} & $0.688\pm0.209$ \\  
	\bottomrule[1.6pt] 
\end{longtable}
%%%%%%
\subsection{Prediction of observables in the NP scenario S3 ($C_9^{\mu, {\rm NP}} = -C_9^{\prime\mu, {\rm NP}} = -1.01$) }
\begin{longtable}{@{}lcrr@{}}
	\toprule[1.6pt] 
	\hspace{0mm} Bin \hspace{10mm} & $P_1$\hspace{22mm} & $P_2$\hspace{24mm} & $P_3$\hspace{10mm}  \\ 
	\midrule 
	$ [0.1,1] $ \hspace{10mm}  & $0.058\pm 0.058$\hspace{14mm} & $0.123\pm 0.004$ \hspace{14mm}  & $ 0.001\pm 0.029 $ \\ 
	$ [1,2] $ & $0.195\pm 0.057$\hspace{14mm} & $0.404\pm 0.011$ \hspace{14mm} & $ 0.005\pm 0.029$ \\ 
	$ [2,4] $ & $0.165\pm 0.058$\hspace{14mm} &  $0.338\pm 0.026$ \hspace{14mm} & $ 0.007\pm 0.027 $ \\ 
	$ [4,6] $ & $-0.031\pm 0.061$\hspace{14mm} &  $-0.068\pm 0.034$ \hspace{14mm}  & $ 0.005\pm 0.026$ \\ 
	$ [1,6] $ & $0.084 \pm 0.060$\hspace{14mm} & $0.172\pm 0.028$ \hspace{14mm} & $ 0.006\pm 0.027 $ \\ 
	\midrule[1.6pt] 
	\hspace{0mm} Bin & $P_4^\prime$\hspace{22mm} & $P_5^\prime$\hspace{24mm} & $P_6^\prime$\hspace{10mm}  \\ 
	\midrule 
	$ [0.1,1] $ & $-0.265\pm 0.032$\hspace{14mm} & $0.921\pm 0.119$\hspace{14mm} & $0.044\pm 0.051$ \\ 
	$ [1,2] $ & $0.124\pm 0.022$\hspace{14mm} & $0.689\pm 0.091$\hspace{14mm} & $0.048\pm 0.033$ \\ 
	$ [2,4] $ & $0.694\pm 0.087$\hspace{14mm} & $0.198\pm 0.053$\hspace{14mm} & $0.044\pm 0.021$ \hspace{1mm}\\ 
	$ [4,6] $ & $0.993\pm 0.119$\hspace{14mm} & $-0.258\pm 0.056$\hspace{14mm} & $0.030\pm 0.040$ \\ 
	$ [1,6] $ & $0.704\pm 0.087$\hspace{14mm} & $0.103\pm 0.048$\hspace{14mm} & $0.039\pm 0.022$ \\  
	\midrule[1.6pt] 
	\hspace{0mm} Bin & $BR~(10^{-7})$\hspace{22mm} & $A_{FB}$\hspace{24mm} & $F_L $\hspace{10mm}  \\ 
	\midrule 
	$ [0.1,1]$& $0.189\pm0.093$\hspace{14mm} & $0.105\pm0.025$\hspace{14mm} & $0.261\pm0.181$ \\ 
	$ [1,2] $ & $0.083\pm0.042$\hspace{14mm} & $0.262\pm0.130$\hspace{14mm} & $0.559\pm0.226$ \\ 
	$ [2,4] $ & $0.146\pm0.080$\hspace{14mm} & $0.179\pm0.112$\hspace{14mm} & $0.653\pm0.218$ \\ 
	$ [4,6] $ & $0.169\pm0.088$\hspace{14mm} & $-0.041\pm0.033$\hspace{14mm} & $0.599\pm0.225$ \\ 
	$ [1,6] $ & $0.398\pm0.208$\hspace{14mm} & $0.102\pm0.061$\hspace{14mm} & $0.610\pm0.223$ \\  
	\bottomrule[1.6pt] 
\end{longtable}
%%%%
\subsection{Prediction of $R_{K_2^\ast}$ }
\begin{longtable}{l l l l l}
	\toprule[1.6pt] 
	\hspace{3mm} Bin &\hspace{6mm} SM \hspace{5mm} &\hspace{5mm} S1\hspace{5mm} & \hspace{5mm}S2\hspace{5mm} & \hspace{15mm}S3 \hspace{5mm}\\ 
	\midrule
	$ [0.1,1] $ \hspace{5mm} & $0.986\pm 0.005$ \hspace{5mm}& $0.943\pm 0.061$ \hspace{5mm}& $0.910\pm 0.060$ \hspace{5mm}&\hspace{5mm} $0.96\pm 0.049$ \\
	$ [1,2] $\hspace{5mm} &  $0.998\pm 0.003$ \hspace{5mm} & $0.918\pm 0.098$\hspace{5mm}  & $0.829\pm 0.068$ & \hspace{5mm} $0.960\pm 0.092$ \\ 
	$ [2,4] $ &  $0.996\pm 0.002$  & $0.861\pm 0.071$ & $0.749\pm 0.020$ &\hspace{5mm} $0.903  \pm 0.068$ \\  
	$ [4,6] $ &  $0.996\pm 0.002$  & $0.813\pm 0.028$  & $0.714\pm 0.008$ & \hspace{5mm} $0.847\pm 0.025$ \\  
	$ [1,6] $ & $0.996\pm 0.002$  & $0.851\pm 0.056$  & $0.749\pm 0.017$ & \hspace{5mm} $0.888 \pm 0.052$ \\    
	\bottomrule[1.6pt] 
\end{longtable}
%%

%%%%%%%%%%%%%%%%%%%%%%%%%%%%%%%%%%%%%%%%%%%%%%%%%%%%%%%%%%%%%%%%%%%%%%%%%%%%%%%%%%%%
%%%%%%%%%%%%%%%%%%%%%%%%%%%%%%%%%%%%%%%%%%%%%%%%%%%%
% ---- Bibliography ----
%
\bibliography{refs.bib}
\bibliographystyle{utcaps_mod.bst}
\end{document}